\documentclass[longauth]{aa} 
\usepackage{graphicx}
\usepackage{txfonts}
\usepackage{natbib}
\bibpunct{(}{)}{;}{a}{}{,} 
\begin{document}
\newcommand{\smst}{{\it sample ${\cal M}$ST}} 
\newcommand{\smcor}{{\it sample ${\cal M}$COR}} 
\newcommand{\st}{{\it ${\cal M}$ST}} 
\newcommand{\mcor}{{\it ${\cal M}$COR}} 
\newcommand{\masq}{magn \arcsec$^{-2}$}
\newcommand{\Jykms}{Jy\,km\,s$^{-1}$}
\newcommand{\mjyb}{mJy beam$^{-1}$}
\newcommand{\HO}{$H_0$}
\newcommand{\kms}{km\,s$^{-1}$}
\newcommand{\acm}{cm$^{-2}$}
\newcommand{\teff}{$T_{eff}$} 
\newcommand{\kmsmpc}{km\,s$^{-1}$\,Mpc$^{-1}$}
\newcommand{\mjb}{mJy beam$^{-1}$}
\newcommand{\jb}{Jy beam$^{-1}$}
\newcommand{\mjbc}{mJy beam$^{-1}$ channel$^{-1}$}
\newcommand{\jykms}{Jy km s$^{-1}$}
\newcommand{\logm}{log($\cal M/M_\odot$)} %
\newcommand{\msun}{$\cal M_\odot$}
\newcommand{\msunyr}{$\cal M_\odot$\,yr$^{-1}$}
\newcommand{\lsun}{{$L_{\odot,B$}}}
\newcommand{\lb}{{L$_{\rm B}$}}
\newcommand{\mlsun}{{\rm M}$_\odot/{\rm L}_{\rm B_{\odot}$}}
\newcommand{\kmsMpc}{km s$^{-1}$ Mpc$^{-1}$}
\newcommand{\hi}{H{\small I}}
\newcommand{\MHI}{{$\cal M}_{\rm HI}$}
\newcommand{\m}{\hbox{$^{\rm m}$}}
\newcommand{\s}{\hbox{$^{\rm s}$}}
\newcommand{\h}{\hbox{$^{\rm h}$}}
\newcommand{\dn}{D$_{\rm n}$4000}
\newcommand{\halpha}{H$\alpha$}
\newcommand{\ewhalpha}{EW(H$\alpha$)}
\newcommand{\hbeta}{H$\beta$}
\newcommand{\ewhbeta}{EW(H$\beta$)}
\newcommand{\hgamma}{H$\gamma$}
\newcommand{\hepsilon}{H$\epsilon$}
\newcommand{\hdelta}{H$\delta$}
\newcommand{\hdeltaA}{H$\delta_{A}$}
\newcommand{\ewhdelta}{EW(H$\delta$)}
\newcommand{\hds}{H$\delta$~strong}
\newcommand{\oii}{[O{\small II}]}
\newcommand{\ewoii}{EW[O{\small II}]}
\newcommand{\oiii}{[O{\small III}]}
\newcommand{\ewoiii}{EW[O{\small III}]}
\newcommand{\neiii}{[Ne{\small III}]} 
\newcommand{\caii}{Ca{\small II}} 
\newcommand{\ie}{{i.e.\,}}
\newcommand{\eg}{{e.g.\,}}
\newcommand{\psb}{{post-starburst}}
\newcommand{\Psb}{{Post-starburst}}
\newcommand{\ka}{{$k+a$}}
\newcommand{\Ka}{{$K+a$}}
\newcommand{\chandra}{{\it Chandra\/}}
\newcommand{\xmm}{{XMM-{\it Newton\/}}}
\newcommand{\cgs}{{${\rm erg~cm}^{-2}~{\rm s}^{-1}$}}
\newcommand{\lum}{{\rm erg~s$^{-1}$}}

   \title{{K+a galaxies in the zCOSMOS Survey: \thanks{Based on data obtained
   with the European Southern Observatory Very Large Telescope, Paranal,
   Chile, program 175.A-0839.}}}

   \subtitle{Physical properties of systems in their post-starburst phase}
   
   \titlerunning{K+a Galaxies in zCOSMOS}

\author{
  D. Vergani  \inst{1,2}                    
\and G.~Zamorani \inst{2}
\and S.~Lilly \inst{3} 
\and F.~Lamareille \inst{4} 
\and C.~Halliday \inst{5} 
\and M.~Scodeggio \inst{6} 
\and C.~Vignali \inst{2} 
\and P.~Ciliegi \inst{2} 
\and M.~Bolzonella \inst{2} 
\and M.~Bondi \inst{7}  
\and K.~Kova\v{c} \inst{3} 
\and C.~Knobel \inst{3}
\and E.~Zucca \inst{1}
\and K.~Caputi \inst{3}
\and L.~Pozzetti \inst{1} 
\and S.~Bardelli \inst{1} 
\and M.~Mignoli \inst{1} 
\and A.~ Iovino \inst{8}
\and C. M.~Carollo \inst{2}
\and T.~Contini \inst{4}
\and J.-P.~Kneib \inst{9}
\and O.~ Le F\`{e}vre \inst{9}
\and V.~Mainieri \inst{10}
\and A.~Renzini \inst{11}
\and A.~Bongiorno \inst{12}
\and G.~Coppa \inst{2}
\and O.~Cucciati \inst{9}
\and S.~de la Torre \inst{8}
\and L.~ de Ravel \inst{9}
\and P.~Franzetti \inst{6}
\and B.~Garilli \inst{6}
\and P.~Kampczyk \inst{2}
\and J.-F.~Le Borgne \inst{4}
\and V.~Le Brun \inst{9}
\and C.~Maier \inst{2}
\and R.~ Pello \inst{4}
\and Y.~ Peng \inst{2}
\and E.~ Perez Montero \inst{4}
\and E.~ Ricciardelli \inst{11}
\and J.~D.~Silverman \inst{2}
\and M.~ Tanaka \inst{10}
\and L.~ Tasca \inst{6}
\and L.~ Tresse \inst{9}
\and U.~Abbas \inst{13}
\and D.~ Bottini \inst{6}
\and A.~ Cappi \inst{1}
\and P.~ Cassata \inst{9}
\and A.~ Cimatti \inst{2}
\and L.~ Guzzo \inst{8}
\and A.M.~Koekemoer \inst{14}   
\and A.~ Leauthaud\inst{15}    
\and D.~ Maccagni \inst{6}  
\and C.~ Marinoni \inst{16}    
\and H.J.~ McCracken \inst{17} 
\and P.~ Memeo \inst{6}
\and B.~ Meneux \inst{12,18}   
\and P.~ Oesch \inst{2}
\and C.~ Porciani \inst{19}    
\and R.~ Scaramella \inst{20}  
%
\and P.~Capak\inst{21}
\and D.~Sanders\inst{22}
\and N.~Scoville\inst{21}
\and Y.~Taniguchi\inst{23}
}

   \offprints{\mbox{D. Vergani}, \email{daniela.vergani@oabo.inaf.it}}

 \institute{INAF-Osservatorio Astronomico di Bologna, Via Ranzani 1, I-40127, Bologna,
   \email{daniela.vergani@oabo.inaf.it} 
\and 
{Universit\`a di Bologna, Dipartimento di Astronomia, Via Ranzani 1, I-40127, Bologna, Italy} \and
{Institute of Astronomy, Swiss Federal Institute of Technology (ETH H\"onggerberg), CH-8093, Z\"urich, Switzerland.} \and
{Laboratoire d'Astrophysique de Toulouse-Tarbes, Universit\'{e} de Toulouse, CNRS, 14 avenue Edouard Belin, F-31400 Toulouse, France} \and
{INAF - Osservatorio Astrofisico di Arcetri, Largo Enrico Fermi 5, I-50125 Firenze, Italy}\and
{INAF - IASF Milano, Milan, Italy} \and
{INAF - Istituto di Radioastronomia, Via Gobetti 101, I-40129 Bologna, Italy}\and
{INAF Osservatorio Astronomico di Brera, Milan, Italy} \and
{Laboratoire d'Astrophysique de Marseille, Marseille, France} \and
{European Southern Observatory, Karl-Schwarzschild-Strasse 2, Garching, D-85748, Germany} \and
{Dipartimento di Astronomia, Universita di Padova, Padova, Italy} \and
{Max-Planck-Institut f\"ur extraterrestrische Physik, D-84571 Garching, Germany} \and
{INAF - Osservatorio Astronomico di Torino, 10025 Pino Torinese, Italy}\and
{Space Telescope Science Institute, 3700 San Martin Drive, Baltimore, MD 21218}\and
{LBNL \& BCCP, University of California, Berkeley, CA 94720, USA}\and
{Centre de Physique Theorique, Marseille, Marseille, France} \and
{Institut d'Astrophysique de Paris, UMR 7095 CNRS, Universit\'e Pierre et Marie Curie, 98 bis Boulevard Arago, F-75014 Paris, France.} \and
{Universitats-Sternwarte, Scheinerstrasse 1, D-81679 Muenchen, Germany} \and
{Argelander-Institut f\"ur Astronomie, Auf dem H\"ugel 71, D-53121 Bonn, Germany}\and
{INAF, Osservatorio di Roma, Monteporzio Catone (RM), Italy} \and
{California Institute of Technology, MC 105-24, 1200 East California Boulevard, Pasadena, CA 91125, USA}\and
{Institute for Astronomy, University of Hawaii, 2680 Woodlawn Drive, Honolulu, HI, 96822}\and
{Research Center for Space and Cosmic Evolution, Ehime University, Bunkyo-cho, Matsuyama 790-8577, Japan}
}
 \date{Received 30 June 2009; accepted 9 September 2009}

  \abstract {}{ The identities of the main processes triggering and quenching
    star-formation in galaxies remain unclear. A key stage in evolution,
    however, appears to be represented by post-starburst galaxies. To
    investigate the prevalence of these galaxies and their impact on galaxy
    evolution in general, we initiated a multiwavelength study of galaxies
    with \ka\ spectral features in the well-studied COSMOS field.}  { We
    examine a mass-selected sample of \ka\ galaxies in the COSMOS field at
    $z=0.48-1.2$ using the spectroscopic zCOSMOS sample. To classify galaxies
    in their \ka\ phase, we use a spectroscopic criterion, based on the
    amplitude of the \hdelta\ absorption line and the absence of the \oii\
    emission line. We develop our analysis for a well-defined sample by
    imposing stringent confidence levels on the spectroscopic redshifts and
    spectral measurements.  We compare our results for two mass-selected
    samples of star-forming and quiescent galaxies selected using a purely
    spectral classification scheme from the 10,000 zCOSMOS catalogue (\ie,
    based on measurements of 4000~\AA\ break and \ewoii).}
{In our mass-limited sample, \ka\ galaxies occupy the brightest tail
  of the luminosity distribution. They are as massive as quiescent
  galaxies and populate the green valley in the colour versus
  luminosity (or stellar mass) distribution. A small percentage
  ($<8$\%) of these galaxies have radio and/or X-ray counterparts
  (implying an upper limit to the SFR of $\sim 8$\msunyr). Over the
  entire redshift range explored, the class of \psb\ galaxies is
  morphologically a heterogeneous population with a similar incidence
  of bulge-dominated and disky galaxies. This distribution does not
  vary with the strength of the \hdelta\ absorption line but instead with
  stellar mass in a way reminiscent of the well-known mass-morphology
  relation.  The results about the incidence of asymmetries and the
  concentration of the light distribution derived from HST/ACS images
  imply that this galaxy population possibly represents an intermediate
  stage of galaxy evolution. 
Although \ka\ galaxies are also found in underdense regions, they
appear to reside typically in a similarly rich environment as
quiescent galaxies on a physical scale of $\sim 2-8$~Mpc, and in
groups they show a morphological early-to-late type ratio similar to
the quiescent galaxy class.
With the current data set, we do not find evidence of statistical
significant evolution in either the number/mass density of
\ka\ galaxies at intermediate redshift with respect to the local
values, or the spectral properties, although more solid results on this
and other aspects will be obtained following the completion of the
survey.}
{Several mechanisms related and unrelated to the environment are at
  work in quenching star-formation activity in galaxies on short
  timescales ($<1$~Gyr). Those galaxies, which are affected by a sudden
  quenching of their star-formation activity, may increase the stellar
  mass of the red-sequence by up to a non-negligible level of $\sim
  10$\%.}
 \authorrunning{D. Vergani
  et al.}  \titlerunning{Post-starburst galaxies in COSMOS field} 

\maketitle

\section{Introduction}
\label{sec:intro}

Numerous studies have demonstrated that a bimodal population of
galaxies exists up to high redshift \citep[e.g.,][]{franzetti07,
  mignoli09}. The first population consists of galaxies that are
actively star-forming, gas-rich, rotationally-supported, and
morphologically disk-dominated. The second population represents
typically quiescent galaxies that are gas-poor, pressure-supported,
and have spheroidal morphologies. As noted by many studies
\citep[e.g.,][]{bow98, be00}, observational evidence exists that many
galaxies have experienced morphological transformations at various
redshifts. \citet{be00} reported that a decline has occurred in the
mass density of irregular galaxies between {\it z}~$\simeq$~1 and
today, in contrast to an increase for regular galaxies. Further
support of this redistribution between galaxy types has been provided
by study of the stellar mass function by many different surveys, e.g.,
K20 \citep{fon04}, COMBO-17 \citep{bor06}, DEEP2 \citep{bun06},
SWIRE-VVDS-CFHTLS \citep{arn07}, SDSS \citep{mar07}, and VVDS
\citep{ver08}.  It is important to assess the role of environment in
these transformations of galaxy type.  For instance, it has been shown
that galaxies in dense regions have modified distributions of \hi, the
primary source of fuel for star-formation, and truncated
star-formation activity \citep[e.g.,][]{gio85, cay90, koo04}.
However, several other mechanisms may also contribute (see
below). Thus, important questions in galaxy formation and evolution
are whether star-forming galaxies are the precursors of the quiescent
galaxy population observed at the present day, or whether there are
alternative paths in galaxy formation, and, if so, what are the
primary processes involved and how are they controlled?

The main limiting factor in answering this question is that we still have no
true understanding of what triggers star formation in galaxies and why this
activity is suppressed in some of them. There is a class of galaxies that
might shed light on this issue. They are called $k+a$ or $E+A$ galaxies
(depending on the author), or simply galaxies in the post-starburst phase. In
this work, we adopt the spectral terminology $k+a$ instead of the
morphological $E+A$ nomenclature.

Post-starburst galaxies, discovered by Dressler and Gunn (1983), exhibit
peculiar spectral characteristics, i.e., strong Balmer absorption lines,
indicative of an intense star-formation epoch in the past billion years, and
an absence of emission lines, a signature of a lack of ongoing
star-formation. On the basis of these spectral signatures, they have been
assumed to represent galaxies in a transitional stage between being a
star-forming, late-type galaxy and a passive, early-type system.

Various mechanisms have been proposed to explain the characteristic
spectral features of \psb\ galaxies. Cluster-related mechanisms, such
as ram-pressure gas stripping, harassment, or strangulation
\citep{gunn72, larson80, balogh00} are efficient in suppressing the
star-formation activity of a galaxy.  Events such as galaxy mergers
and interactions \citep{toomre72, barnes92}, commonly found in both
the field and in clusters (although more efficient in the field), have
been proposed. The large fraction of morphological irregularities (of
similar physical timescale as a \psb\ phase) and galaxy companions in
the vicinity of these galaxies, support the hypothesis that either a
dynamical interaction or a merger with exchange of material may have
suppressed star formation.  Strong AGN/SN feedback might also
contribute to star-formation quenching \citep{springel05, hopkins07},
depending on the stellar mass \citep{kaviraj07}.

This galaxy population was first discovered, and has been primarily
studied since, inside clusters \citep{dre83,cou87,tran03,pog99,pog09},
even if \psb\ galaxies have also been found in the field.  Numerous
studies have been undertaken to study the properties of field
\psb\ galaxies at low redshift \citep[][]{liu95, zab96, cha01, nor01,
  got03, qui04, yang04}, and at $z>0.3$ \citep{ham97, dre99, pog99,
  bal99, tran04, yang08, pog09}.  These studies commonly concluded
that \psb\ galaxies are a heterogeneous parent population.

\Psb\ galaxies appear to be rare at any epoch and discrepant results
about their nature are found in the literature.  The identifiable
sources of discordance are numerous. Besides differences in selection
criteria and spectral resolution, the main source of discrepancy may
be the different signal-to-noise ratio achieved at different
redshifts. This implies that we should also investigate the dependency
of the underlying properties on the host galaxies (e.g., the sampling
of only the most massive or brightest systems).  Furthermore, because
of the observational constraints, the majority of studies cover small
areas and might probe specific types of environments (large-scale
structures in pencil surveys, or studies in clusters).
In addition to the observational limitations, finding post-starburst
galaxies is also challenging because of the short duration of the
post-starburst phase (1--1.5~Gyr) (see \citet{cou87} and
\citet{bar96}). Because they represent a small fraction of the total
galaxy population, the systematic effects caused by the aforementioned
problems require particular attention when interpreting the results in
a general framework.

In this paper, we present for the first time a mass-selected sample of
post-starburst galaxies at intermediate redshift in a wide variety of
different environments. This mass-selected sample and its parent galaxies are
extracted from the largest multiwavelength survey existing up to now, the
Cosmic Evolution Survey, or COSMOS \citep{sco07} imaged with HST/ACS
\citep{anton07}.  We provide an overview of the main physical properties of
these galaxies using data of unprecedented quality and completeness in this
study.  We present a first attempt to quantify their cosmic evolution. A more
detailed study of this subject and its relation to the general picture of
galaxy evolution over the past 8~Gyr will be presented on completion of
the survey.

The present work is organized as follows: the data and sample
selection are presented in Sect.\,2; results are described in
Sect.\,3, and a summary is given in Sect. 4. Throughout this work, we
assume a standard cosmological model with $\Omega_M = 0.3$,
$\Omega_\Lambda = 0.7$, and $H_0 = 70 \, \mathrm{km} \,
\mathrm{s}^{-1} \, \mathrm{Mpc}^{-1}$.  Magnitudes are given in the AB
system.

\section{Sample selection}
\label{sec:sample}

\subsection{The spectroscopic survey: zCOSMOS}

The zCOSMOS project \citep{lil07} is a redshift survey of galaxies in the
COSMOS field consisting of two observing programs.  The {\it ``deep part''}
focuses on high-$z$ galaxies ($1.4 < z < 3.0$) using a combined flux-limited
($B_{AB} < 25.25$) and colour selection (these observations are not used in
the present study).  The {\it ``bright part''} of zCOSMOS project aims to
acquire spectra of about 20,000 galaxies ($I_{AB} < 22.5$) with a rather
uniform sampling rate (about 60--70\%) across the 1.7~deg$^2$ COSMOS field, a
superb spectroscopic redshift success rate ($>97$\% at $0.5<z<0.8$), and high
velocity accuracy ($\sim$~100\,\kms).  The {\it bright part} is observed with
the red $R\sim 600$ VIMOS MR grism covering the spectral range 5500--9500\AA\
(2.55\AA\ dispersion), which provides an ideal configuration for detecting key
spectral features in galaxies out to $z\sim1.2$.  This paper deals with the
first ``10k'' galaxy data set completed to date for the {\it bright part} of
the zCOSMOS survey, which we refer to hereafter as the zCOSMOS survey.  All
details of data acquisition, reduction, redshift measurements, and their
quality can be found in \cite{lil07} and \citet{lil09}, which also describes
other important aspects of this survey.

\begin{figure}[t!]
\begin{center}
\resizebox{\hsize}{!}{\includegraphics[width=8cm,angle=0]{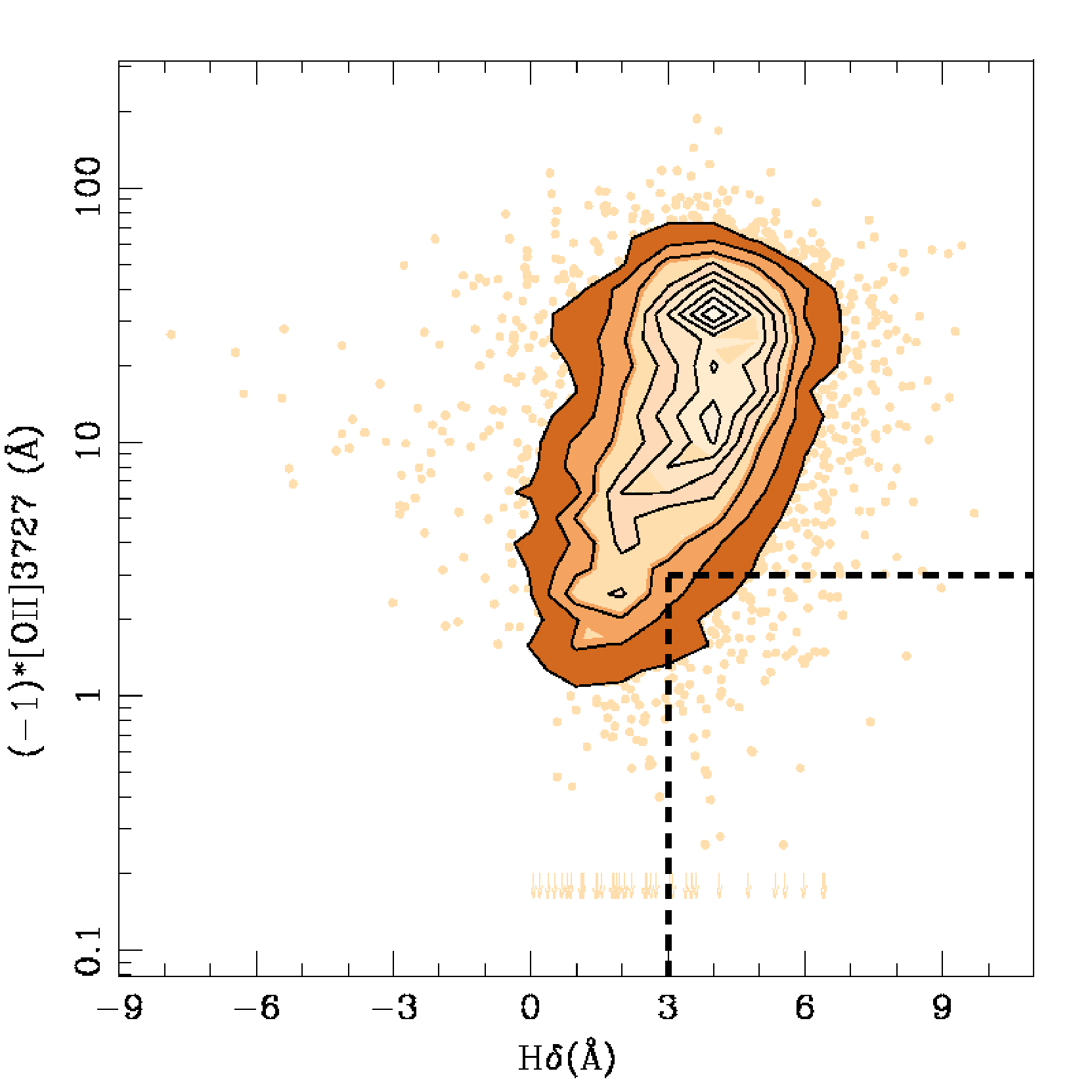}} 
\resizebox{\hsize}{!}{\includegraphics[width=6cm,angle=-90]{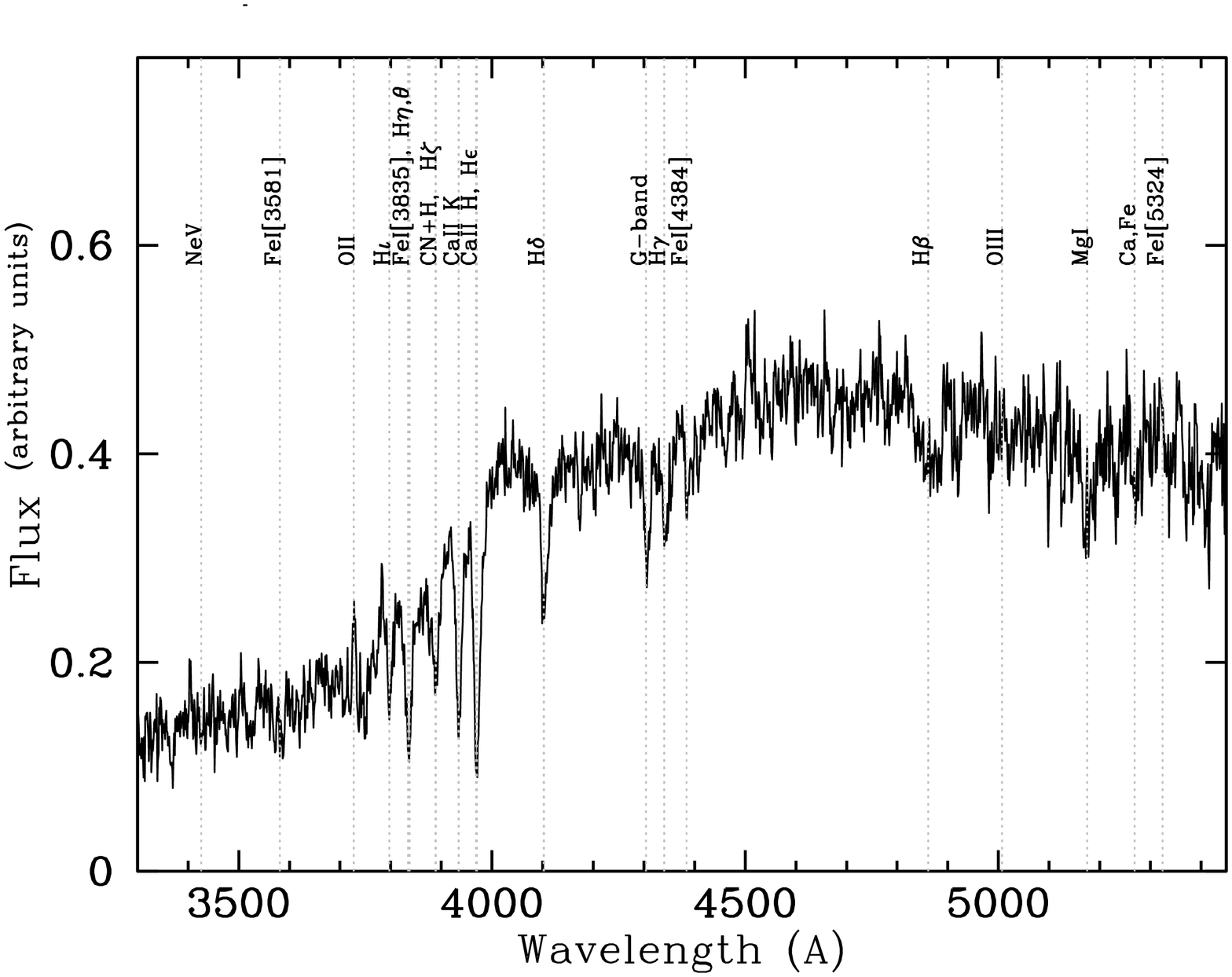}}
\end{center}
\caption{({\it Top}) Distribution of equivalent widths of \hdelta\
  vs. EW\oii\ for all galaxies in the 10k-parent sample in the
  redshift interval $0.48 \le z \le 1.2$.  This plot shows the
  criteria adopted to select the post-starburst candidate galaxies,
  i.e., the region between \hdeltaA$>+3$\AA\ and EW\oii$>-3$\AA\
  marked with dashed lines.  Isophotal contours are in steps of 10\%
  with the faintest isophote level starting at the 10\% level. ({\it
    Bottom}) Template spectrum of a typical galaxy in a post-starburst
  phase obtained by stacking 69 zCOSMOS post-starburst galaxies more
  massive than \logm$ > 10.0$ (corrected by selection effects as
  explained in Sect.\,2.5).}
\label{fig:sel_tem} \end{figure}

\subsection{The spectral measurements and sample selection}
\label{sec:spec}

We measure the spectral features in our 10k sample with an automated
software package (platefit\_{VIMOS}). Here we provide a brief outline
of the fitting procedure. We refer to \cite{lam06b} for full details,
and both \citet[]{tre04} and \cite{bri04} for further details of the
original pipeline developed for the higher resolution SDSS
spectra. The best-fit stellar continuum derived from a grid of stellar
population synthesis models (Bruzual \& Charlot 2003, hereafter BC03)
was subtracted from the observed spectrum.  Any remaining residual was
removed by fitting a low-order polynomial to the continuum-subtracted
spectrum, and all emission lines were then fitted simultaneously with
a Gaussian profile. Finally, absorption features and spectral breaks
were measured after subtracting emission lines from the original
spectrum.

In this paper, we use measurements of the spectral index \hdeltaA\ as
defined by \cite{wor97} and \cite{bal99}, the equivalent width of the
\oii$\lambda$3727 doublet (\ewoii), the equivalent width of the
\oiii$\lambda$5007 (\ewoiii), and the amplitude of the
4000~\AA\ break (\dn) \citep{bal99}. A careful treatment of
spectral fitting is mandatory because both emission and absorption
lines are present in the Balmer lines. The index \hdeltaA\ and the
equivalent width of the lines are defined such that positive values
refer to absorption features.

Authors usually adopt \hdeltaA\ and \ewoii\ features in their criteria
for selecting post-starburst galaxies, and/or \halpha\ depending on
the spectral coverage, although some exceptions exist. For example,
\citet{tran04} used a combination of \hdelta\ and \hgamma\ indices, in
addition to \oii, \citet{yan06} added an empirical limit to
\ewhbeta\ in constraining \ewoii, and some others used, in addition to
a threshold on emission lines, the relative abundance of A-type to
K-type stars, i.e., the ratio A/K \citep[e.g.,][]{qui04}.  In this
paper, we adopt a conventional definition using both the
\oii\ emission line and \hdeltaA, to be able to compare with the
majority of other works, although any comparison is difficult and only
possible qualitatively. In a future paper, based on the 20k galaxy
sample, we will examine the effects of different selection criteria on
a larger sample of \hdelta-enhanced galaxies \citep[\eg,][]{leb06} and
the possible contribution from obscured AGNs \citep{yan06}. In the
present paper, our post-starburst candidate criteria are that the
equivalent width of \hdeltaA\ is larger than $+3$\AA\ and the
equivalent width of the emission line \oii\ is larger than $-3$\AA.
This is illustrated in the top panel of Fig.\,\ref{fig:sel_tem}, where
the equivalent width of \hdeltaA\ is plotted versus \ewoii\ for the
10k-parent galaxies in the redshift interval $0.48 \le z \le 1.2$ (the
redshift bin explored in this work, which is imposed by our
instrumental set-up). This plot shows the spectroscopic criteria
adopted to select the post-starburst candidate galaxies. As is evident
in Fig.\,\ref{fig:sel_tem}, the region corresponding to our
\ka\ candidates is not completely separate from that of the 10k-parent
population as it is in the SDSS data (cf. the left panel of Fig.\,2 by
\cite{yan08}). This is mainly due to the errors associated with the
spectral measurements. To evaluate their impact on the classification
scheme, we tested different empirical thresholds and applied a
confidence level to the spectral features when defining our
samples. This approach aims to exclude (include) residual signatures
of ongoing star formation and to test the reliability of our spectral
measurements. While the numbers of objects change depending on the
thresholds adopted, of course, we obtained an optimal compromise
between the ability to complete a robust, statistical analysis and
reliable results by including only galaxies with \ewoii\ larger than
-3\AA\ and \ewhdelta\ larger than 3\AA\ at the 2$\sigma$ confidence
level.

To maintain strong control on occasional poor wavelength calibration
that can affect in particular the bluest part of our spectra, we
visually inspected each spectrum of potential candidates. For spectra
with doubtful measurements of \ewoii, we imposed an additional limit
on the equivalent width of the \ewoiii\ line. We used a subsample of
high signal-to-noise 10k spectra and high confidence spectral
measurements in which both \ewoii\ and \ewoiii\ lines were detectable
in the spectra, to define this limit on the \oiii\ lines. From the
best-fit model solution, we inferred a threshold for the \ewoiii\ of
$-6.9$\AA.  Furthermore, from all of our catalogues, we removed both
stars and broad-line AGNs and adopted a high-quality flag scheme for
the spectroscopic redshifts ($>95.5\%$ confidence level) (for details
see \citet{lil07} and \citet{lil09}).  Although we inspected each
spectrum individually, our confidence level in the redshift
measurements is 100\% for our \psb\ sample.  We also performed quality
checks of our post-starburst candidate spectra to verify the impact of
residual sky lines on spectral features. We examined the reduced VIMOS
spectrum for the impact of fringing effects, which may be significant
at wavelengths above $\lambda$~$>$~8,500\AA\ because of the thinned,
back-illuminated CCDs used in VIMOS \citep{sco05}.

The final sample presented in Sect.\,\ref{sec:results} contains
spectra that passed all of the aforementioned tests. 

Using the constraints described above, we obtain a sample of 74
post-starburst galaxies in the flux-limited zCOSMOS bright sample in
the redshift interval $z\in[0.48-1.2]$. Three of them are at redshifts
higher than $z>1$.  For one object, the computation of a precise
absolute magnitude is impossible because of the lack of sufficient
photometric coverage, and it is excluded from our statistical analysis
(for details see in Zucca et al. (2009)).

\subsection{Post-starburst spectral template}
\label{sec:template}

Given the difficulties in selecting this rare population of galaxies,
we provide a template spectrum\footnote{This template can be accessed
  from the web page {\tt
    http://www.bo.astro.it/$^{\sim}$daniela/template.fits}.} to the
community of the \psb\ galaxy population obtained by stacking 69
zCOSMOS \psb\ galaxies more massive of \logm$ > 10.0$ (corrected by
selection effects as explained in Sect.\,2.5). A description of the
steps performed to obtain this template spectrum is given in the
following text.

The bottom panel of Fig.\,\ref{fig:sel_tem} shows the template
spectrum, which exhibits the distinctive spectral features of galaxies
that have recently experienced a starburst before its star-formation
was quenched prior to observations.  These features can be summarized
as follows:

{ i}) The strength of the \oii\ nebular emission line, which is
emitted in H{\small II} regions around O and B stars, is almost absent
(EW\oii$<-$2.5\AA) in our template spectrum. Given that \oii\ is an
indicator of recent star formation (with lifetime of 10$^7$~yr), this
property indicates that the star formation during the epoch of the
observation has stopped.  The \halpha\ emission line is another robust
indicator of ongoing star formation, but cannot be detected with our
adopted $R\sim 600$ VIMOS MR-grism observational setup, which covers
the spectral range $5500\AA < \lambda < 9500\AA$.\\
{ ii}) Another typical spectral feature of this galaxy population
is the strong absorption features of Balmer lines. Although
\halpha\ and \hbeta\ are often filled in by nebular emission that can
make their interpretation difficult, \hdelta\ can be used to estimate
the stellar ages.  From late-type stars to earlier-type stars, the
\hdelta\ strengthens. This can be achieved if A-stars dominate the
main-sequence contribution near the epoch of the observation.\\
{ iii)} As noted originally by \cite{rose84}, the ratio of the
equivalent widths of the \caii~H$\lambda3968+$\hepsilon\ to
\caii~K$\lambda3934$\AA\ spectral lines is constant in stars later
than about F02, but increases dramatically for earlier type stars as
the \caii\ lines weaken and \hepsilon\ strengthens. This spectral
property is also illustrated in our template spectrum, where the
\caii~H$\lambda3968+$\hepsilon\ is deeper than the
\caii~K$\lambda3934$\AA\ absorption lines.\\
{ iv)} We note the lack of emission from \oiii\ and \hbeta\ lines in
our template spectrum, which indicates that narrow-line Seyferts do
not represent an important component of our sample, in contrast to
other works \citep[e.g.,][]{yan08}.

The combined use of the two SFR estimators (\hdelta\ and \oii), which are
sensitive to stars of different stellar masses and lifetimes, allows us to
obtain in the next section a quite precise cosmic clock of their recent
($<1$~Gyr) star-formation history, and to characterize their physical
properties.

\begin{figure}[t]
\begin{center}
{\includegraphics[width=9.5cm,angle=0]{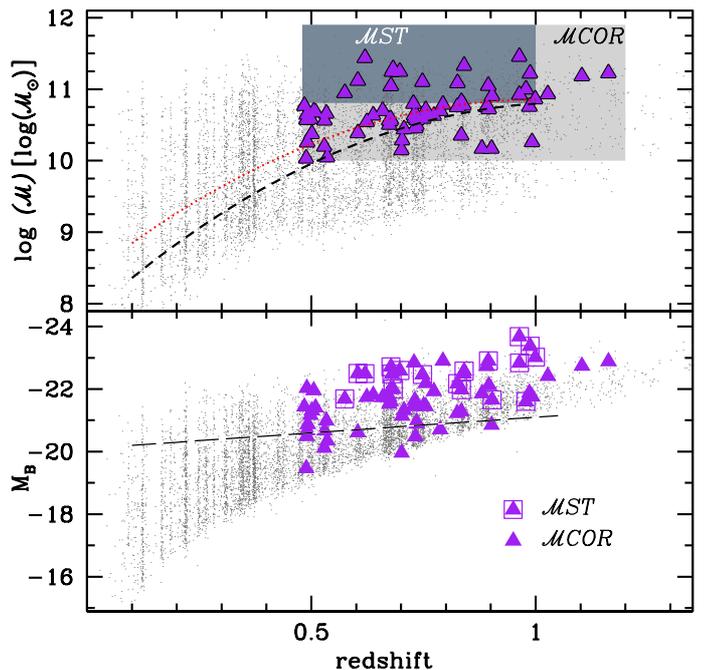}}
\end{center}
\caption{ Stellar mass and B-band absolute magnitude as a function of
  redshift of the zCOSMOS bright sample (points, gray symbols). The
  (violet) triangles show the location of \psb\ galaxies. The light
  and dark gray areas show the selection in stellar mass adopted in
  \smcor\ and \smst, respectively. The dashed (black) line in the
  upper panel illustrates the completeness in mass-to-light ratio for
  the global population (for comparison, it is plotted with a dotted,
  red line the completeness limit for the early-type population). The
  dashed (black) line in the bottom panel is plotted as reference to
  illustrate that our galaxies of \smst\ are brighter than the
  evolving-luminosity limit at M$_B <-20.1 -
  z$.} \label{fig:mass1} \end{figure}

\subsection{Stellar masses}

Since we studied galaxies that are understood to have experienced a
recent secondary burst of star formation, we used Charlot \& Bruzual
models with the inclusion of secondary bursts (Charlot \& Bruzual in
prep., or CB07). These models have been frequently used to model the
complex star-formation history of a galaxy more realistically than a
purely exponentially declining law \citep{kau03a, bri04, sal05}.

Stellar masses were derived by fitting stellar population synthesis
models from the library of CB07 to the broad-band optical (CFHT: u, i,
Ks; Subaru: B, V, g, r, i, z; Capak et al.  2007) and near-infrared
(Spitzer/IRAC: 3.6$\mu$m, 4.5$\mu$m; Sanders et al. 2007) photometry
by minimizing the chi-square test statistic for each galaxy. The
estimate of the stellar mass provided by the best-fit model spectral
energy distribution is the current mass contained in stars in each
galaxy at the epoch of the observation obtained by integrating the
star formation history over the galaxy age and subtracting the
cumulative mass lost during stellar evolution.

The measurement of stellar mass included some assumptions. We used a
Chabrier initial mass function \citep{cha03} with a lower and upper
cutoff at 0.1 and 100\msun, a dust attenuation described by the model
of \citet{cf00}, and metallicities between 0.1 and 2 $Z$\sun. The
star-formation history was modelled by two components. The first
component was a continuous star-formation model with an exponentially
declining law of the form SFR(t) $\propto$ exp (-t/$\tau$), where
timescale $\tau$ and age t are in the range $\tau/$[Gyr] $=$[1,$\infty$] and
t/[Gyr] $=[0.1,20]$. The second component was represented by random
bursts superimposed on the first component that had an equal
probability of occurring throughout the life of the galaxy.

Of course the results depend on the adopted assumptions and models
used. For example, the assumption of a Chabrier initial mass function
produces a consistent difference in the values of the stellar mass at
fixed age that is within a factor of 1.7, but with a small dispersion,
of those derived with the Salpeter IMF prescription \citep{poz07}. The
inherent uncertainty in the initial mass function adopted is not a
global limitation for results based on stellar masses given the
constancy of this rigid shift over a wide range of star formation
histories, although to limit the various degeneracies a large set of
grids of stellar population synthesis models were used. Further
details of the stellar mass measurements can be found in \citet{bol09}
and \citet{poz07, poz09}.

\subsection{\Psb\ representative sample}
\label{sec:weight}

Since our original selection is based on a flux-limited survey, we
used subsamples of galaxies complete in terms of stellar mass
(adopting two independent approaches) to avoid selection biases.

In the first approach, we adopted a conservative limit to the stellar
mass of \logm$ > 10.8$ in the redshift bin $z\in[0.48-1.0]$ (from now
on, this stellar Mass STringent criterion refers to \smst). This
stringent mass limit was defined using Mock catalogues from the
Millennium simulation \citep{delucia07} of the zCOSMOS bright sample
(see \cite{men09} for an exhaustive explanation on the subject). We
selected 18 \psb\ galaxies for this conservative mass-complete sample,
or \smst.

To maximize the quality of the statistical analysis, we also
constructed a second sample, or \smcor, including galaxies more
massive than \logm$ > 10.0$ in the redshift bin $z\in[0.48-1.2]$, and
accounting for under-represented galaxies at high $z$ by correcting
with statistical weights and a $V_{tot}/V_{max}$ formalism
\citep{sch68}.  The total volume sampled by the survey within the
redshift bin explored is denoted by $V_{tot}$.  The volume $V_{max}$
is the maximum volume in which each galaxy of a given $I_{AB}$
magnitude is still observable.
We applied the statistical weights adopted in the studies of
luminosity and mass functions of the zCOSMOS sample \citep{zuc09,
  bol09, poz09}. These weights are the product of the inverse of the
target sampling rate (TSR) and the inverse of the spectroscopic
sampling rate (SSR). The target sampling rate is the ratio of the
objects spectroscopically observed to the total number of objects in
the parent photometric catalogue. The spectroscopic sampling rate is
the ratio of the number of observed objects with reliable redshifts to
the total number of observed objects, calculated in bins of apparent
magnitude.  The apparent magnitude dependence takes into account the
decline towards fainter magnitudes, in our ability to measure a
redshift. A more complex scheme that introduces a redshift dependence
of SSR does not alter the final results appreciably in the redshift
bin explored here (see Bolzonella et al. 2009). In the \smcor, the
mass completeness limit, which depends on both redshift and
mass-to-light ratio, also takes account of the colour-magnitude
relation \citep{poz09, bol09}.

The second approach permits us to increase the number of objects (from
18 to 69) in our analysis and explore a wider redshift bin
($z\in[0.48-1.2]$) than that of $z\in[0.48-1.0]$ for \smst.
However, we always compare the results obtained with this MCOR sample
with those obtained with the complete MST sample to validate
further our conclusions.

Figure\,\ref{fig:mass1} shows the stellar mass of the zCOSMOS bright
sample as a function of redshift and, overplotted, the data of the
\psb\ galaxies in our mass-complete samples (with violet, triangle
symbols).  We note that the conservative limit imposed by Mock
catalogues agrees well with the completeness in the observed
mass-to-light ratio for the global population. This limit is similar
to the more conservative limit obtained for the early-type population
(defined using the spectrophotometric classification described in
Zucca et al. 2009). The statistical weights and mass completeness
adopted are the values used in the luminosity and mass function
studies by \cite{zuc09}, \cite{bol09}, and \cite{poz09}. In the same
figure, we also plot the B-band absolute magnitude as a function of
redshift. It is important to note that our \smst, which is complete in
terms of stellar mass such that \logm$ > 10.8$, is brighter than the
evolving-luminosity limit at M$_B <-20.1 - z$ (illustrated with the
dashed, black line overplotted in the left panel of
Fig.\,\ref{fig:mass1}).

\subsection{Comparison samples}

We compiled two comparison catalogues of strongly star-forming and
quiescent galaxies. For these galaxies, we also adopted consistently
the same stellar mass completeness criteria, redshift flags, and
confidence levels on the spectral measurements as used in defining the
class of \psb\ galaxies.

The star-forming and quiescent galaxy classes were selected using a
pure spectral classification. We combined measurements of the
amplitude of the 4000~\AA\ break, taken as an age estimator of
the underlying stellar population, and the equivalent width of the
\oii$\lambda$3727 line, assumed to be an indicator of the
star-formation activity.  Thus, quiescent galaxies were defined to be
galaxies with a zero, or negligible, star-formation rate inferred
using the \ewoii, i.e., \ewoii$>-3$\AA, and an old underlying
stellar population described by their large 4000\AA\ break
(\dn$>1.5$) (e.g., \citet{kau03b}).  We defined a sample of strong
\ewoii\ emitters by constructing a catalogue of objects with
\ewoii$<-15$\AA\ and small 4000~\AA\ break (\dn$\le1.5$). From
the 10k sample, we also compiled the 10k-parent catalogue of galaxies
following the selection in stellar mass adopted for \smcor\ and \smst,
and all other criteria listed above (from now on when we use the
terminology ``all objects'', we refer to the 10k-parent catalogue).

\begin{figure}[t]
\begin{center}
\includegraphics[width=8cm,angle=0]{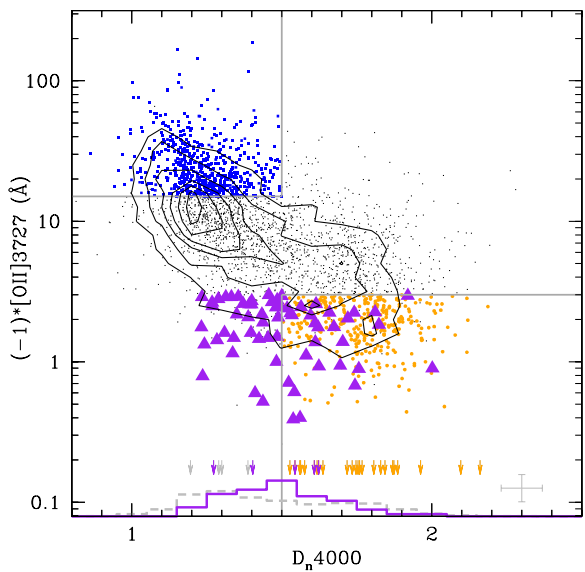}
\end{center}
\caption{ The amplitude of 4000~\AA\ break is plotted as a function of
  \ewoii\ to show the selection criteria used to classify the classes
  of star-forming (square, blue symbols) and quiescent (circle, orange
  symbols) galaxies. Triangle (violet-coded) symbols show the
  \psb\ galaxies. Small points (in black) show the 10k-parent
  catalogue (isophotal contours are in steps of 15\% with the faintest
  isophote level starting at 15\% level). The 4000~\AA\ break
  histograms of the 10k-parent sample (dashed, gray line) and of the
  \psb\ galaxies (solid, violet line) are also shown along the x-axis.
  The vertical lines indicate the location of the cut in \dn\ (at
  1.5). The horizontal lines indicate the location of the two extreme
  thresholds at -3\AA\ and -15\AA\ in \ewoii. Arrows are upper limits
  to \ewoii\ values. Typical errors in the spectral measurements are
  plotted as errorbars at the bottom of the
  panel.} \label{fig:comp} \end{figure}

Our selected samples are displayed in Fig.\,\ref{fig:comp}, where the
selection criteria used to classify galaxies in our comparative analysis are
shown.  This figure illustrates how the three classes of galaxies (highly
star-forming, quiescent, and \psb\ galaxies) populate the plane of the
amplitude of 4000~\AA\ break and \ewoii. Small points and isophotal
contours in steps of 15\%, with the faintest isophote level starting at 15\%,
show the 10k-parent galaxies level.  Using criteria adopted for \smcor, we
obtain a comparison sample of 295 quiescent galaxies and 174 star-forming
galaxies, while we have 178 quiescent galaxies and 45 star-forming galaxies in
\smst. The 10k-parent galaxies number 1508 following the criteria of \smcor,
and 665 for \smst.

\section{Results}
\label{sec:results}

We present the main physical properties of the mass-selected
\psb\ galaxies identified in the zCOSMOS survey in attempting to
obtain a coherent picture of their origin in the context of galaxy
formation. Our analysis refers to galaxies of the statistically
corrected, mass-complete sample at $z\in[0.48-1.2]$ (\smcor) and the
conservative mass-limited sample at $z\in[0.48-1.0]$ ({\smst}).

\subsection{Robustness of our sample: dusty starburst hypothesis?}
\label{sec:hypo}

Galaxies with spectral characteristics similar to those used to define
a post-starburst galaxy are often interpreted to be a system in which
star formation has suddenly and almost completely stopped within the
past few Gyrs.  This cessation of star-formation is assumed to explain
the lack of emission lines in the galaxy spectrum. 

An alternative explanation, however, is that these galaxies do have
emission, but that this is extinguished by a significant amount of
dust.  Galaxies with these characteristics should be dusty, starburst
galaxies (and misclassified \psb\ objects).  This is a hypothesis that
should be verified especially when dealing with galaxies understood to
have experienced a recent, strong burst of star formation.  OB stars
responsible for the \oii\ emission, as other younger stellar
generations, are more strongly obscured by dust than older stellar
populations. This age selective dust extinction enables the true
strength of the Balmer absorption lines to be measured, but only a
fraction of the emission to be detected in stronger starburst galaxies
\citep{pog01,pog09}. To ascertain whether our sample of \psb\ galaxies
may be contaminated by dusty starburst galaxies, we performed
different tests. First, we explored the dusty starburst scenario by
computing both the extinction values and star-formation rate by SED
fitting. Secondly, taking advantage of the available multiwavelength
observations, we explored the X-ray and 1.4~GHz radio data sets
\citep{sch07, cap09, bru09}.

\smallskip \noindent {\bf 1) Dust extinction and SED fitting SFR.}  We
found that \psb\ galaxies have in general low values of extinction as
estimated by fitting CB07 model SEDs (median value $<\tau_{\rm
  V}>=0.6$). The median SFR is 0.6\msunyr, and the 25\% and 75\%
quartiles of the distribution range from 0.01\msunyr\ to 4.2\msunyr.
The distributions of both $\tau_{\rm V}$ and SFRs for \psb\ galaxies
differ at a high confidence level from the same distributions for
star-forming galaxies ($<\tau_{\rm V}>\sim 2$ and $<{\rm
  SFR}>\sim21$\msun). While \psb\ galaxies exhibit a similar
cumulative $\tau_{\rm V}$ distribution to quiescent galaxies, the
latter galaxies have an even lower rate of star-formation activity, as
expected when using the used spectroscopic criteria described in
Sect.\,2 to identify them ($<\tau_{\rm V}>\sim 0.4$ and $<{\rm
  SFR}>\sim0.07$\msun).  For only one \psb\ galaxy do we infer from
SED fitting a high rate of star-formation activity ($\sim 140$\msunyr)
and a high extinction ($\tau_{\rm V}\sim2$). This object is one of the
two \psb\ galaxies detected at both X-ray and radio 1.4~GHz
frequencies, even if the radio detection is only at the $3\sigma$
confidence level (see below).

\smallskip \noindent {\bf 2) 1.4~GHz radio frequency.}  We
investigated the dusty starburst hypothesis further by
cross-correlating our sample with the 1.4GHz VLA-COSMOS survey
\citep{sch07,bon08}. If it is not produced by AGN emission, the radio
flux is indicative of the current star-formation activity (over short
timescales of 10$^8$~yr) unbiased by dust \citep[see][ for a
  review]{con92}.

We found that one \psb\ galaxy has a counterpart at 1.4~GHz radio
frequency with a detection confidence level of $7.5\sigma$ and five
others have possible 1.4~GHz emission at $\simeq 3\sigma$ confidence
level. For the entire sample of 74 \psb\ galaxies, we therefore have a
detection rate of 1.3$\pm$1.3\%, if we adopt the threshold of
4$\sigma$ for assuming a radio detection to be true (as adopted in
\citet{sch07} and \cite{bon08}). This detection rate increases to
8.1$\pm$2.7\%, if we adopt the less conservative threshold of
3$\sigma$ for radio detection.

To investigate the mean radio properties of the \psb\ galaxies, we
performed a stacking analysis of the 1.4~GHz radio images.  From the
stacking analysis, we excluded the six \psb\ galaxies with a possible
radio counterpart plus three additional sources that exhibit, within
2.5\arcsec\ of their optical position, either radio emission more
significant than 3$\sigma$ that is not associated with the
\psb\ galaxy, or a negative radio peak of higher significance than
3$\sigma$.  For the remaining 65 \psb\ galaxies, images of 40 $\times$
40 pixels (corresponding to 14\arcsec\ $\times$ 14\arcsec) were extracted
around the optical position. Finally, all the 65 images were combined
to obtain a median image with a 1$\sigma$ radio noise of $\approx$2
$\mu$Jy.  A marginal detection of $\approx3\sigma$ confidence level
was found at a distance of 1.4\arcsec\ from the centre of the stacked
image with a radio flux of $\approx$7 $\mu$Jy.  To access the
reliability of this detection, we repeated the stacking analysis for 5
different samples, each of them consisting of 75 random positions. The
criterion used to exclude from the stacking, \psb\ galaxies with a
radio detection (or a negative radio peak, see above) was applied
similarly to the simulated samples. All 5 stacked images from the
simulated sample had a 1$\sigma$ radio noise in the range 1.4-1.8
$\mu$Jy. No radio detection was identified for any of the simulated
stacked images.

Therefore, assuming the marginal detection in the stacked image to be
real and assuming its flux of 7 $\mu$Jy to be the mean radio flux of
radio-quiet \psb\ galaxies, we can estimate their mean star-formation
rate, or more realistically its upper limit since some objects may be
AGNs.  We obtained an upper limit of 8.2\msun~yr$^{-1}$ by adopting
the calibration by \citet{bel03}.  The radio luminosity L$_{\rm
  1.4GHz}$ was calculated by assuming a median redshift of
$z_{median}={\sc 0.7014}$.

We then explored the physical properties of the six ($>3\sigma$)
radio-detected \psb\ galaxies.  Radio-detected \psb\ galaxies are
slightly more massive and brighter than non-radio-detected
\psb\ galaxies (log${\cal M/M_\odot}_{\rm 1.4GHz} =11.22 \pm
0.09$ versus log~${\cal M/M_\odot}_{\rm no-1.4GHz} = 11.08 \pm
0.05$; M$_{\rm B,1.4GHz}=-22.64 \pm 0.37$ versus M$_{\rm
  B,no-1.4GHz}=-22.44 \pm 0.14$).  \Psb\ galaxies with a counterpart
at 1.4~GHz also show slightly bluer colours ((U-B)$_{\rm
  1.4GHz}=1.10\pm0.1$) than non-radio-detected \psb\ galaxies
((U-B)$_{\rm no-1.4GHz}=1.18\pm0.03$).  Although these differences are
not statistically significant per se, their properties and low
detection rate suggest that radio-detected \psb\ galaxies might
represent the initial phase in the evolution of \psb\ galaxies, close
to the peak of the starburst (the timescale of radio emission is an
order of magnitude shorter than that of the $k+a$ phase).

\smallskip \noindent {\sc \bf 3) X-ray counterparts.}  To characterize
the multiwavelength properties of \psb\ galaxies, we also searched for
their possible X-ray counterparts using the available data provided by
\xmm\ for the entire 2~deg$^{2}$ of the COSMOS field (Cappelluti et
al. 2009; Brusa et al. 2009) and \chandra\ over the inner
0.9~deg$^{2}$ \citep{elvis09} at considerably deeper X-ray flux
levels. While only one \psb\ galaxy is detected by \xmm\ for a
matching radius of 6\arcsec, five X-ray sources (four ``new'' plus the
\xmm\ one) are detected by \chandra\ (within 4\arcsec\ because of the
sharper PSF) at a detection likelihood $>$10.8 in at least one X-ray
band (see Elvis et al. 2009 and Puccetti et al. 2009 for further
details) among the 54 \psb\ galaxies located in the field of
\chandra\ mosaic.  Visual inspection of all of these sources supports
their likely physical association with the \psb\ galaxies under
investigation.  All of the five \psb\ galaxies detected by
\chandra\ are in regions of moderate-to-high exposure time
($>$100~ks), where more than one \chandra\ pointing is available
(Elvis et al. 2009). However, the limited number of counts
($\approx$~9--40) in the \hbox{0.5--7~keV} band prevents us from
performing an adequate X-ray spectral analysis.  X-ray counts were
converted into fluxes (using a power-law model with photon index
$\Gamma=1.4$, as adopted in the \chandra-COSMOS catalog; Elvis et
al. 2009), and then into 0.5--10~keV rest-frame luminosities (given
the source spectroscopic redshifts of 0.67--1.10) of
1.1--8.6$\times10^{42}$~\lum.

At least for the three sources with the highest X-ray counts in the
hard band (above 2~keV), this luminosity range suggests a significant
contribution from AGN emission.  The assumption of a photon index of
$\Gamma=1.9$, which is typical of unobscured AGN emission, would
decrease X-ray fluxes and luminosities by 25\%.  The optical spectra
of these three sources with the highest X-ray counts in the hard band,
might indicate that they are AGN host galaxies.

Of these five X-ray sources, two have a clear association with sub-mJy
sources in the VLA catalog \citep{bon08}, one of which has the highest
SFR of all sources and dust extinction inferred from SED fitting.

To place constraints on the properties of the X-ray undetected
\psb\ galaxies, X-ray stacking analysis was applied to the remaining
49 sources, extracting source counts from circles of 5\arcsec\ radius
and background counts from a region at least 10\arcsec\ away from the
source position. X-ray sources present in the \chandra\ source catalog
(Elvis et al. 2009) were carefully masked.  To limit the problems with
the PSF at large off-axis angles, we extracted source counts using
only observations where the source position was located within
8\arcmin\ of the field centre (see \cite{miyaji08} and \cite{kim09}
for a description of the tool adopted here).  We did not find any
significant signal from the stacked \psb\ galaxies using \chandra; the
3$\sigma$ flux upper limit to the stacked 0.5--8~keV counts is
$\approx2\times10^{-16}$~cgs, corresponding to a rest-frame 0.5--8~keV
luminosity of $\approx4.9\times10^{41}$~\lum\ at the average redshift
($\langle\ z \rangle \approx$~0.73) of the stacked sources.  This
upper limit to the stacked signal is consistent with emission from
normal and starburst galaxies.

\citet{dew00} reported a detection of an active nucleus in one
post-starburst galaxy selected during a search for ultrasoft sources
in a ROSAT campaign, but no radio emission was detected within a
radius of 15\arcsec\ in the NRAO/VLA and FIRST sky surveys. In
previous studies, similar counter-checks led to a detection rate of
the order of 1-15\%. \citet{miller02} using the galaxies selected by
\citet{zab96}, detected 2 in 15 \psb\ systems in radio continuum
(corresponding to a star-formation rate of about {\sc 2-6} \msunyr),
and 1 in 56 objects were detected by \citet{blak04} (at 7 \msunyr).
Using Chandra observations of the Extended Growth Strip from DEEP2,
\citet{georgakakis08} found an association between \psb\ galaxies at
$z \sim 0.8$ and AGN, supporting a moderate levels of AGN obscuration.
Although these results are difficult to compare because of different
selection criteria and different mass and luminosity ranges, these
results have been used to claim that at least some ``so-called
post-starburst'' galaxies are still actively forming stars, implying
that a selection based of the blue-part of the optical spectra is
imprecise. Some authors have shown that optical spectra typical of
a \psb\ galaxy in the blue part of their spectrum, have substantial
\halpha\ emission \citep[e.g.,][]{liu95}. We fully agree with the
statement by e.g., \citet{miller02} that the minimum requirement for
defining a system in their $k+a$, or \psb\ phase is the coverage of
both blue- and red-parts of the optical spectrum, or complementing the
optical data with other wavelength data sets.  \medskip

Based on our complementary radio-continuum and deep X-ray data, we
found that a negligible fraction of our \psb\ galaxies, selected by
using blue diagnostics of the optical spectra, have possible residual
star-formation activity, and a small percentage shows a contribution
from AGN emission. Thus, the main conclusion of our present study is
that we have selected a robust sample of \psb\ galaxies.

\begin{figure*}
\begin{center}
\hspace*{-1.1cm}
{\includegraphics[width=19.9cm,angle=0]{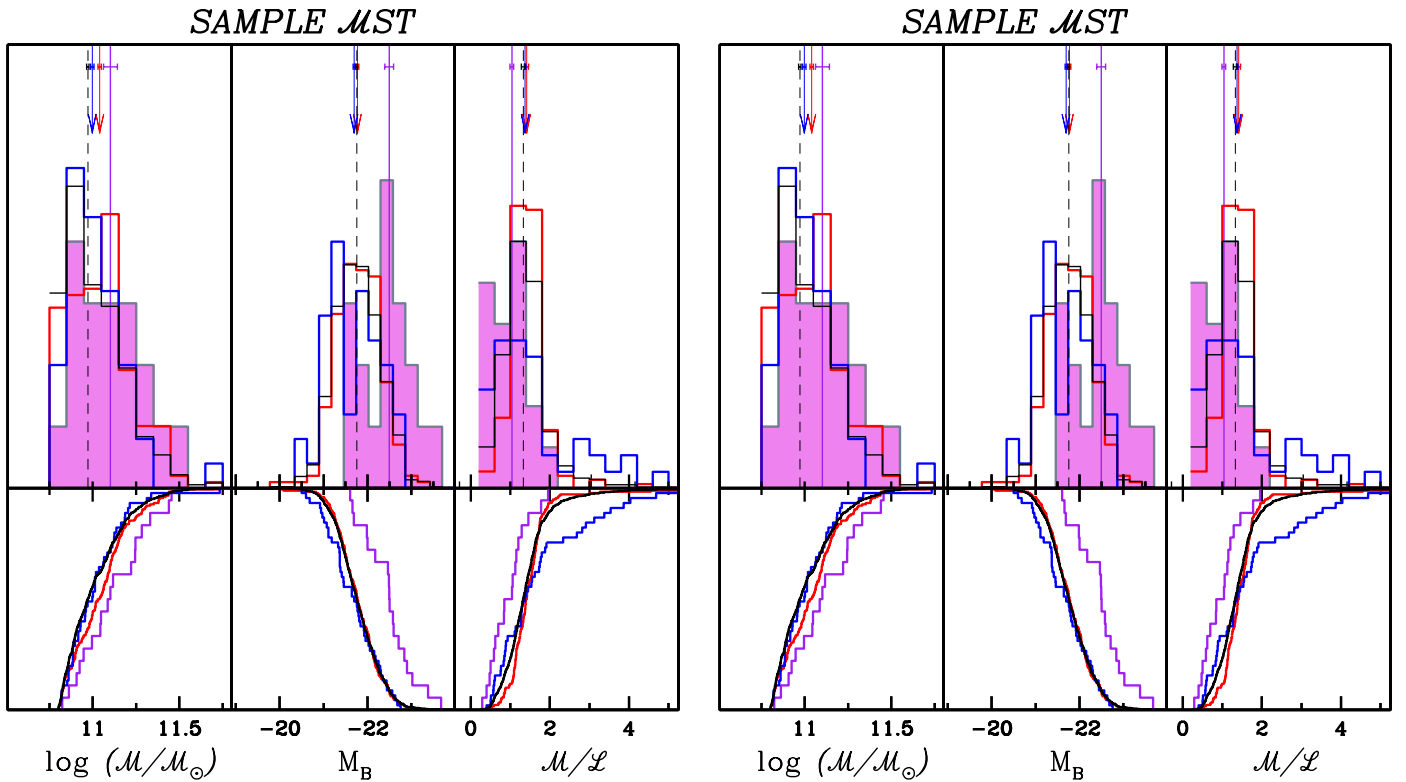}}
\end{center}
\caption{Stellar mass, B-band absolute magnitude, and
  stellar mass-to-light ratio histograms are shown for each class of
  galaxies. The cumulative distributions of each sample are also
  plotted. Median values of \psb\ galaxies and 10k-parent sample are
  plotted as solid and dashed lines, respectively. Blue and red arrows
  are median values of star-forming and quiescent galaxies. The bars
  assigned to each median value correspond to $1\sigma$ error computed
  with resampling technique.  Galaxies classified as post-starburst
  (violet), quiescent (red), star-forming (blue), and the 10k-parent
  (black) sample follow the definition described in Sect.\,2.5.}
\label{fig:mass2} \end{figure*} 

\subsection{Basic properties: luminosities, stellar masses, and colours}
\label{sec:basic}

The B-band absolute magnitude distribution plotted in
Fig.\,\ref{fig:mass2} shows that \psb\ galaxies of the conservative
\smst\ populate the brightest tail of our 10k-parent galaxy
population.  The \psb\ galaxies more massive than \logm$>10.8$ are
brighter by $\sim 0.7-0.8$~mag than the galaxies in both our
comparison classes.  A similar analysis of the \smcor\ galaxies
confirms that the \psb\ galaxy sample is systematically brighter by
$\sim 0.4-0.5$~mag than the median brightnesses compared to median
values of the other galaxies.  Both these results are statistically
significant at more than the $3\sigma$ level. One of the reasons for
this B-band brightness excess is probably a recent burst of star
formation in \psb\ galaxy candidates, as commonly concluded in past
studies.

\Psb\ galaxies also populate the massive tail of the 10k-parent
galaxies (see Fig.\,\ref{fig:mass2}). Their mass distribution differs
from that of the 10k-parent galaxies at $\sim 91$\% confidence level
using the Kolmogorov-Smirnov test \citep{pre92}, and our
\psb\ candidates are as massive as quiescent galaxies.
Our different samples selected using the MCOR criteria show the
typical increasing trend in stellar mass and mass-to-light ratio
expected on the basis of their galaxy classification from star-forming
galaxy population to quiescent galaxies. Given the similar magnitudes
and stellar masses spanned by the star-forming and quiescent galaxies
selected using the MST criteria, this trend is not visible in the
corresponding panels of Fig.\,\ref{fig:mass2}.  \Psb\ galaxies appear
to have a brighter B-band luminosity because of a combination of a
high stellar mass and relatively low mass-to-light ratio.
Among the most massive galaxies (criterion \st), there is a trend for
\psb\ galaxies to be the most massive objects, more massive than
quiescent galaxies (at $1\sigma)$ and star-forming galaxies (at $\sim
2\sigma$).  Since our \psb\ galaxies are already so massive, we may
conclude that their progenitors cannot be {\it common} star-forming
galaxies that have assembled their stellar mass by means of a
continuous, smooth star-formation history and that have stopped, for
whatever reason, forming stars. Indeed to justify the observed stellar
mass assembly of a \psb\ galaxy with a typical stellar mass of \logm$
= 10.58$ (see Fig.\,\ref{fig:mass2}), a specific-SFR (SFR$/{\cal M}$)
of $\sim 1.2 \times 10^{-9}$ (or SFR$\sim 20$~\msunyr) is required for
a galaxy with a stellar mass of \logm$ = 10.23$ (where 10.23 is the
median stellar mass found for our star-forming galaxies). This
specific-SFR value is typical of an object with a high star-formation
efficiency. Furthermore, we can also exclude the dry merging
hypothesis, i.e., the merging of galaxy progenitors in the absence of
gas, as the strength of the \hdelta\ absorption lines implies a recent
burst of new stars, i.e., that fresh gas should have been recently
available.  Therefore, the possible progenitors might be the {\it
  already} massive, less numerous star-forming galaxies at each cosmic
time, or/and the {\it most} actively star-forming sources that in a
short period of cosmic time can assembly sufficient stellar mass: in
other words, the \ka's progenitors might be galaxies with the highest
values of specific star formation rates at any epoch, assuming that
the mass assembly is a process sustained by star formation activity at
the observed rate, i.e., no merger events are invoked.

\begin{figure}[bt!]
\begin{center}
\includegraphics[width=8.0cm,angle=0]{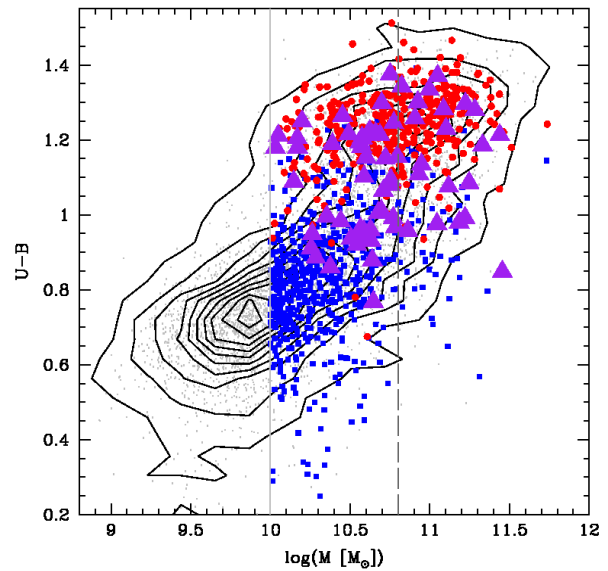}
\end{center}
\caption{ Distribution of the (U-B) colour versus the stellar mass for
  \psb\ galaxies (triangle, violet colour-coded symbols) and
  comparison samples (quiescent galaxies, circle, red-coded symbols,
  and star-forming galaxies, square, blue-coded symbols). Isophotal
  contours are in steps of 10\% with the faintest isophote level
  starting at the 10\% level. The vertical solid (dashed) line
  indicates our mass-completeness at \logm$ > 10.0$ ($> 10.8$).}
  \label{fig:col} \end{figure}

Quiescent and star-forming galaxies populate the red sequence and the
blue cloud, respectively, as shown in Fig.\,\ref{fig:col} for \smcor.
The peak of the distribution for quiescent galaxies is at $(U-B)\sim
1.3$, and star-forming galaxies are distributed around a locus of
$(U-B)\sim 0.8$ in \smcor\ (colours are slightly redder for
star-forming galaxies in \smst\ by $\sim 0.2$~mag, due to the higher
stellar mass selection).  The underpopulated region in-between the two
distributions, i.e., a region called the green valley, is the locus
within which our \psb\ galaxies are located. In the colour-magnitude
diagram, they are on average $(U-B)\sim 0.25$ redder than star-forming
galaxies, and $\sim 0.1$ bluer than quiescent galaxies.
Although the median distribution of colours appears to imply that the
green valley is populated by \psb\ galaxies, they span a wide range of
colours. This property is confirmed in Fig.\,\ref{fig:comp}, where the
4000~\AA\ break histograms of the 10k-parent sample and \psb\ galaxies
are shown along the x-axis (with a solid, violet line and dashed, gray
line, respectively). The 4000~\AA\ break of \psb\ galaxies ranges
between quite small values (typical of star-forming galaxies) and
quite high ones (characteristics of quiescent, passive galaxies), or
$1.2 < D_n4000 < 2$, with a peak at $D_n4000 \sim 1.5$.

Based on this result, selecting \psb-candidate galaxies such as those
that exhibit stronger Balmer absorption lines than expected for their
4000~\AA\ break, e.g., in a way similar to the criterion adopted by
\citet{wild08}, should underestimate their true fraction. Our result
might justify the low detection of this class of galaxies in the VVDS
field. Wild et al. find only 5 post-starburst galaxies in the VVDS
field, scaling from their quoted 18 \hds\ galaxies {\it with} nebular
emission lines to 5 properly-defined \psb\ galaxies (i.e., those
objects among the 18 detected ones {\it without} nebular emission
lines).

Our result indicates that \psb\ galaxies, which populate the green
valley for a very short time, provide a suitable galaxy population to
occupy the gap between blue and red galaxies in the colour-magnitude
relation observed at any redshift from the local Universe up to $z
\sim 1.-1.5$. Although \ka\ galaxies may represent some of the
galaxies in the green valley, we do not however propose that they
constitute the entire population of green-valley galaxies.

\subsection{Morphologies and structural parameters}
\label{sec:morpho}

We investigated the morphological properties of \psb\ galaxies using
the high spatial resolution HST/ACS F814W images
\citep{anton07}. Galaxy morphologies contain the imprints of stellar
populations, gas content, and dynamical structures. Thus, if an event
such as merging has occurred in the recent history of a galaxy, its
signature should be observed, given that the timescale of
gravitational disturbances induced by merging is similar to that of
the \psb\ phase.

We used the ZEST morphological classification of \cite{zest} based on five
non-parametric diagnostics to define early, late, and irregular morphological
classes of all studied galaxies in this work. 
We also explored in more detail properties such as the asymmetry index, the
degree of irregularities, and the concentration of the light.

\begin{figure}[ht!]
\begin{center}
\includegraphics[width=9.3cm,angle=0]{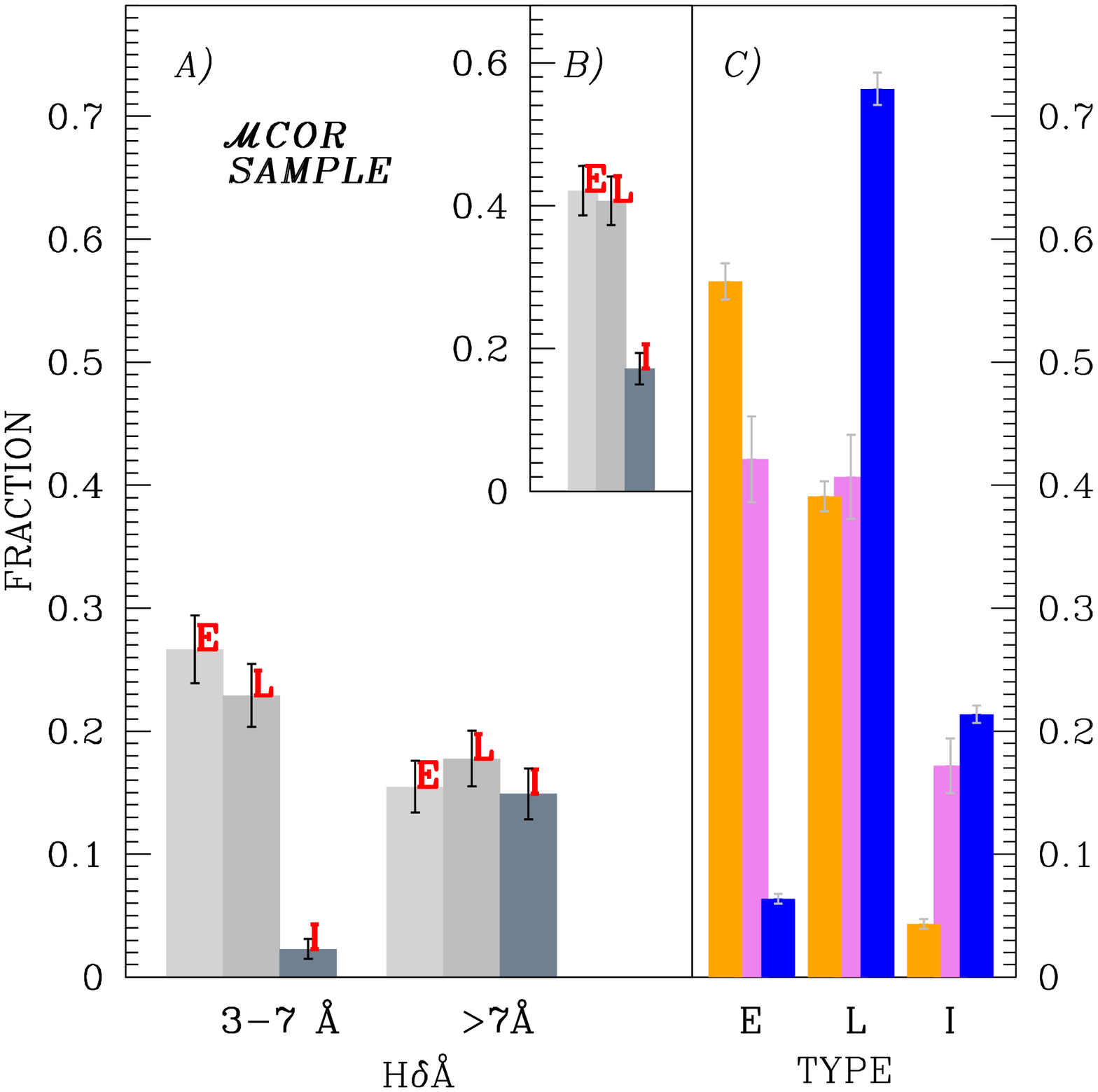}

\includegraphics[width=9.3cm,angle=0]{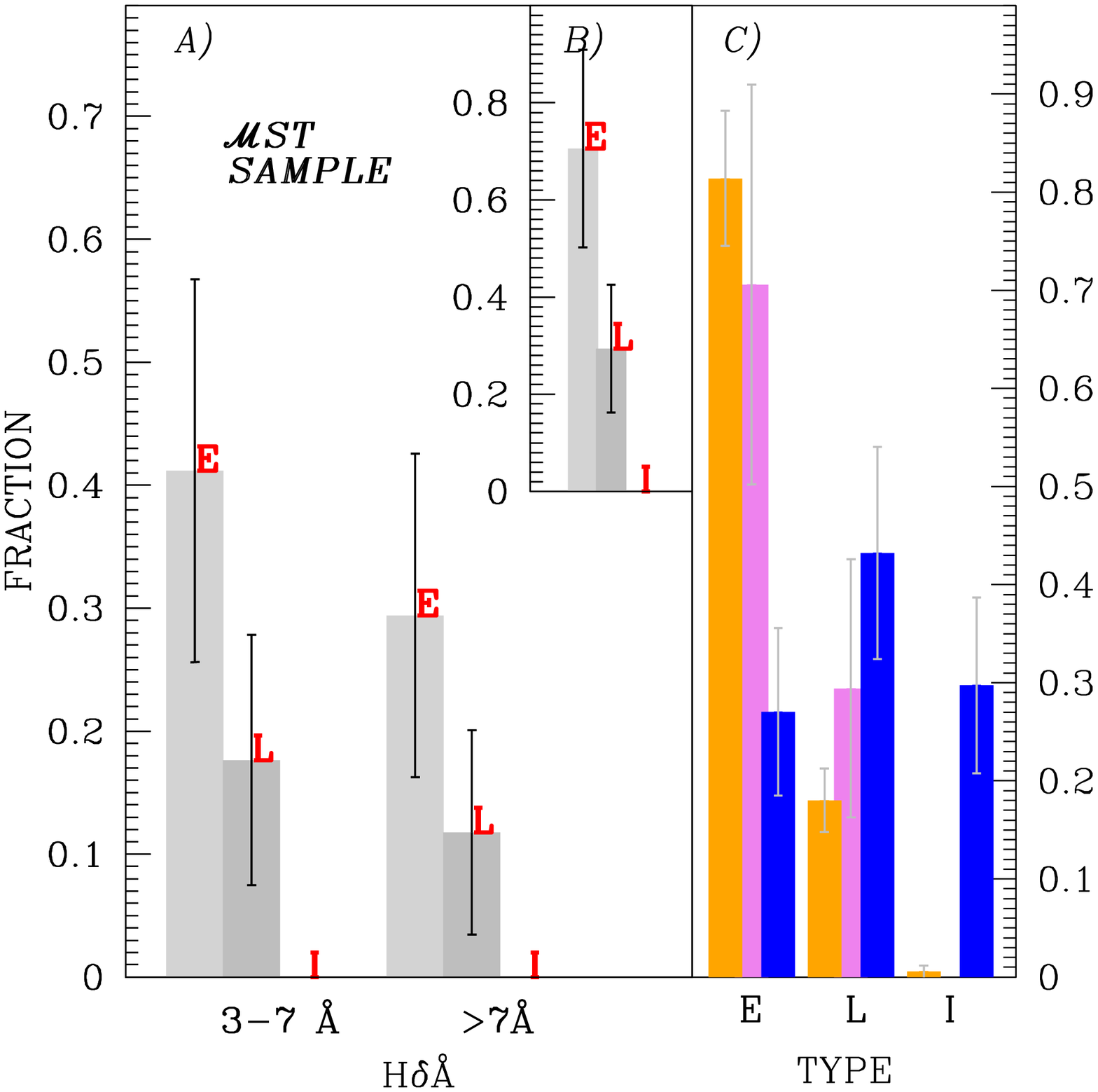}
\end{center}
\caption{Morphological classification of \psb\ galaxies for
  \smcor\ ({top}) and \smst\ ({bottom}). In {\it panel A} we show the
  fraction of the three different morphological types in
  \psb\ galaxies (bulge-dominated (E), disky-dominated (L) and
  irregulars (I)) as a function of the strength of \hdelta. Similar
  plot is shown in {\it panel B} for \psb\ galaxies (independently by
  the \hdelta\ strength). {\it Panel C} shows the fraction of the
  three different morphological types in quiescent galaxies (orange)
  compared to the sample of \psb\ (violet) and star-forming galaxies
  (blue), respectively. The error bars represent the Poissonian
  errors.}
\label{fig:morph} \end{figure}

The upper panel of Fig.\,\ref{fig:morph} illustrates that a similar
fraction of \psb\ galaxies in the sample \mcor\ are bulge-dominated
(E) and disky-dominated (L) (see {panel B}).  How the galaxies are
distributed between the so-called early- and late-type divisions is
independent of their \hdelta\ strength shown in panel~A of
Fig.\,\ref{fig:morph}.  There is, instead, an increase with \hdelta\
in the fraction of \psb\ galaxies in irregular systems. In {panel C},
we show the relative fractions of the three different morphological
types (bulge-dominated, disky-dominated, and irregulars) that are \psb\
galaxies compared to the same fractions for the samples of quiescent
and star-forming galaxies.
As expected, star-forming galaxies are mostly disky-dominated with
$\sim 20-30$\% having irregular morphologies, while the majority of
the quiescent galaxies are bulge-dominated.  Morphologically, \psb\
galaxies are a heterogeneous population, although we note that the
most massive \psb\ galaxies (sample \st) are predominantly
bulge-dominated systems (see lower panels of Fig.\,\ref{fig:morph})
and have a similar distribution among morphological types as quiescent
galaxies. It is interesting to note that in the \smst, there is no
\psb\ galaxy with an irregular morphology, but we also note the low
number statistics in the current \smst.

In some studies, \psb\ galaxies are found to be mostly bulge-dominated
and/or have a morphology that is consistent with being early-type
\citep{got03, qui04, tran04, blak04, got05, bal05, pog09}, and in some
others have a high incidence of disks and spiral arms
\citep[e.g.,][]{fra93, cou94, dre94, cal97, dre99}.  For example,
\citet{qui04} visually inspected a sample of 160 galaxies from the
SDSS at $0.05 < z < 0.20$, and classified 90\% of them as
bulge-dominated systems.
\citet{yang04} used the term ``diverse morphologies'' to describe five
high-resolution HST images of k+a galaxies selected from the sample of
\citet{zab96}, although we note that their selection did favour the
bluest systems, i.e., the ones with smallest \dn\ that have
experienced most recently a significant starburst episode.  Their
detailed study using bulge-to-disk two-dimensional fitting
decomposition of four of them identified one barred S0, two disky
systems, and one unclassifiable galaxy. Their inferred central light
distributions were found to have a profile similar to, but more
luminous, than that of a typical early-type power-law. At higher
redshifts, \citet{tran04} found more bulge-dominated systems in their
$k+a$ galaxy sample than in field galaxies. The sample was selected
from a program designed to study galaxy evolution in clusters, and
$k+a$ galaxies were selected using a conservative constraint on both
the \hdelta\ and \hgamma\ Balmer indices and according to the absence
of \oii\ emission line.  A study based on visual classification of
ACS/HST images by \citet{pog09} identified a higher fraction of S0 and
Sa types among field \psb\ galaxies than \citet{dre99} who instead
found a higher fraction of later types among \psb\ galaxies.
  
Based on the morphological results obtained for the present
\psb\ galaxy sample, we agree with the widely established result that
\psb\ galaxies are morphologically a heterogeneous population. We also
found an increasing fraction of early-type morphologies at increasing
stellar mass in agreement with the well known mass-morphology relation
\citep[e.g.,][]{tasca09}. We might find in the future with larger data
sets spanning a wider range of galaxy properties, that the flux limits
and other selection effects might play a critical role in the
morphological classification, e.g., favouring an early-type assignment
for brighter/massive systems as found in this work.

\begin{figure}[b!]
\begin{center}
\hspace*{-0.7cm}
\includegraphics[width=9.9cm,angle=0]{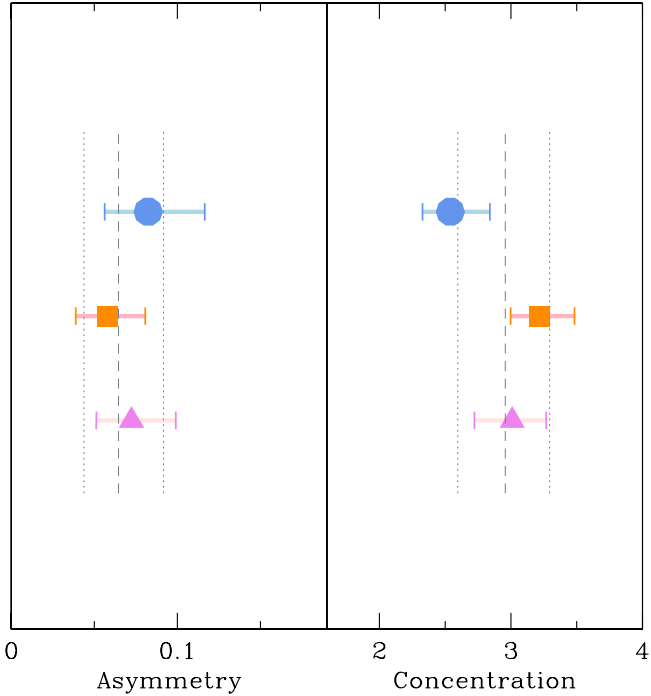}
\end{center}
\caption{The properties of asymmetries and concentrations for each class of
galaxies (\psb\ with triangles, violet-colour coded; quiescent galaxies with
squares, orange colour-coded; and star-forming galaxies with circles,
blue-colour coded).  In both panels, the vertical dotted lines show the median
(dashed) and extreme quartiles (dotted) for the 10k-parent sample.  The
extreme quartiles of each distribution are used to show the dynamical range of
density values in each population. The typical error obtained with resampling
technique is well within the symbols used to plot the median value.}
\label{fig:ac} 
\end{figure}

Apart from being heterogeneous in their morphologies, the second
common feature used to describe \psb\ galaxies by most authors is the
large fraction of asymmetries and clumpiness properties associated
with them (e.g., fine structure, and tidal tails).
We quantify the degree of asymmetry in the light distribution using the
asymmetry coefficient computed as the difference between the image rotated
through 180 degrees and the original frame following the usual prescription
\citep[e.g.,][]{con00} (for details see \citet{sca07}).  Figure\,\ref{fig:ac}
shows that for \psb\ galaxies in the \smcor, the median and range of the
asymmetry index are intermediate between values for quiescent galaxies (and
the 10k-parent sample) and star-forming galaxies.  Comparing the asymmetry
distributions of \psb\ galaxies with those of star-forming and quiescent
galaxies, they both differ at higher than the 99.9\% KS-test confidence level.
Similar results are obtained for the light concentration, which is
defined to be the ratio of the radius including 80\% to that including
20\% of the galaxy light and measures the central density
of the galaxy light distribution.  In Fig.\,\ref{fig:ac}, we show that
the light concentration indices of \psb\ galaxies are typically
in-between the values for quiescent and star-forming
galaxies.
These differences in both the morphological classification and the
parameters of asymmetry and light concentration of \psb\ galaxies and that of
both star-forming and quiescent galaxies, suggest that \psb\ galaxies represent
an intermediate stage of galaxy evolution.

Using the ZEST scheme, the distortion of the light distribution is
quantified by the clumpiness index. This index is available for
galaxies classified as early- and late-types, i.e., not for irregular
galaxies, which by definition are very distorted and for which the
index might be associated with only the maximum degree of clumpiness.
The flags for the clumpiness index range from zero, indicative of a
smooth light distribution, to three, corresponding to a clumpy
light distribution (see Table 2, \citet{zest}).
We found that the majority of quiescent galaxies have a smooth light
distribution (86\% have flags of $0$ or $1$), whereas star-forming
galaxies tend to have a classification consistent with having a clumpy
structure (70.2\% have flags $2$ or $3$). \Psb\ galaxies have a
frequency distribution of concentrations that overall is similar to
that of quiescent galaxies. This is indicative of a smooth light
distribution and consistent with a picture in which the star formation
experienced in the recent past by these galaxies, has ultimately faded
to produce a smoother morphology.

Using Hubble Space Telescope (HST) WFPC2 images, and in agreement with
our results, \citet{caldwell99} detected smoother morphologies in six
\psb\ galaxies than in the starburst galaxies of their sample. In
contrast, the latter galaxies had centrally concentrated star
formation (we note, however, weak \oii\ emission in two of the six
objects that identify them more appropriately with the \hds\ galaxy
class).  \citet{tran04} detected both centrally concentrated light
emission and clumpy regions in the HST WFPC2 images of five \psb\
galaxies.  Although no conclusive results can be drawn because of the
widely ranging properties of \psb\ galaxies, it remains plausible that
they represent a transient phase between an intense
centrally-concentrated burst of star-formation and a more quiescent
stage of star-formation activity.

\subsection{Environmental properties} 
\label{sec:delta}

In this section, we explore the typical environment of \psb\ galaxies,
considering for example whether they live preferentially in overdense regions,
and on which physical scales (from group- to cluster-scales). We also
investigate the relative numbers of group galaxies of particular
morphologies. This aspect of the \psb\ galaxy population, i.e., the study of
their environment, is of particular interest insofar as it represents an
attempt to shed light on the physical mechanisms responsible for quenching the
star formation in this class of galaxies. Apart from the unprecedented quality
and completeness of this data set, the zCOSMOS survey provides an ideal sample
for studying the environmental properties because it is designed to detect
structures down to the physical scales of galaxy groups with a velocity
uncertainty of approximately $\sim 100$~\kms, or $0.00036\times (1+z~)$
\citep[see][]{lil09}.

We used the three-dimensional density estimators derived by \citet{kov09a}
with the ZADE (Zurich Adaptive Density Estimator) algorithm.  We refer to
\citet{kov09a} for a full description of the methodology for which we provide
only a brief outline.
Given the current spectroscopic sampling rate ($\sim 30$\%) of the 10k
sample, we attempted to study the entire volume targeted by zCOSMOS by
adding the 30k photometric redshifts of all COSMOS galaxies to the
presently available 10k spectroscopic redshifts.
In this way using the reconstructed three-dimensional galaxy
distribution, we used the density estimates based on a volume-limited
catalog after application of an adaptive kernel (based in turn on
nearest neighbours). This choice allowed us to account for the
redshift galaxy evolution over the redshift range covered in this
analysis.
We used the nearest neighbour to determine the projected local galaxy
density.  The environment of each galaxy relative to the mean density
at a given redshift ($<\rho(z)>$) was expressed as an overdensity
$\delta$, where $1+\delta=\rho / <\rho(z) >$. The typical errors
associated with this estimator are of the order of ${\rm log}
(1+\delta)=0.1-0.15$ \citep[see Fig.\,4 of][]{kov09a}.  As tracers, we
used galaxies with magnitudes brighter than M$_{\rm B} <$ ($20.5 - z$)
over the redshifts range sampled in this analysis (0.48-1). The
typical comoving scales sampled range between $\sim 2$~Mpc (when we
use the 5$^{\rm th}$ nearest neighbour estimator) and $\sim 8$~Mpc
(using the 20$^{\rm th}$ nearest neighbour estimator). Several
estimates of the overdensity of each galaxy are available, i.e.,
weighting by either stellar mass or luminosity. In this work, however,
we used the number-weighted estimators, i.e., no further weighting was
applied to our environmental estimators.

To investigate the role of the environment on the physical scales of groups of
galaxies, we used the zCOSMOS optically-selected group catalog \citep{kno09}.
Both the algorithm for group detection and the group catalogues were presented
in \citet{kno09}. We describe briefly the main points of the algorithm and
refer the reader for an exhaustive description to the respective paper.  Two
combined group-finding algorithms (the friend-of-friend method and the Voronoi
tesselation density estimator) were implemented to obtain the final catalogue
of galaxy groups. This combined approach, tested with Mock catalogues, lowers
the impact of systematic effects, such as spurious detections and missing
groups, group fragmentation, or overmerging.  Properties such as velocity
dispersion, dynamical mass, and group richness, were available for each group
with an associated error of the order of 20\%. The bright magnitude limit (at
$I_{AB} < 22.5$) of the survey explains in part the relatively small number of
detected group members (the majority of them have 2-4 individual members). The
final catalogue lists a total of approximately 800 groups with a completeness
ranging between 70\% to 90\% \citep[see Fig.\,8 of][]{kno09} and a fraction of
20\% being interlopers.

\begin{figure*}
\begin{center}
\includegraphics[width=8.9cm,angle=0]{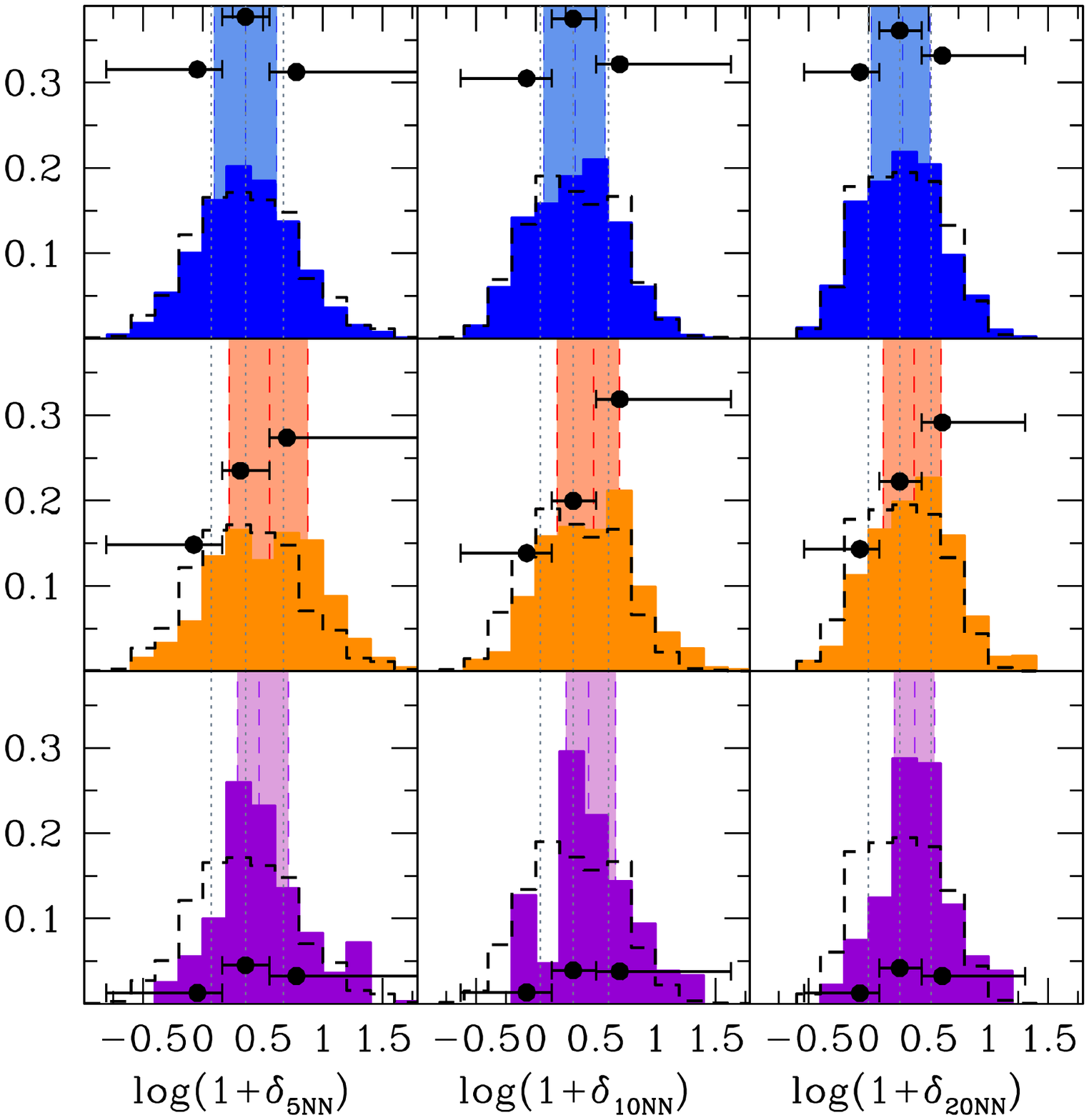}
\includegraphics[width=8.9cm,angle=0]{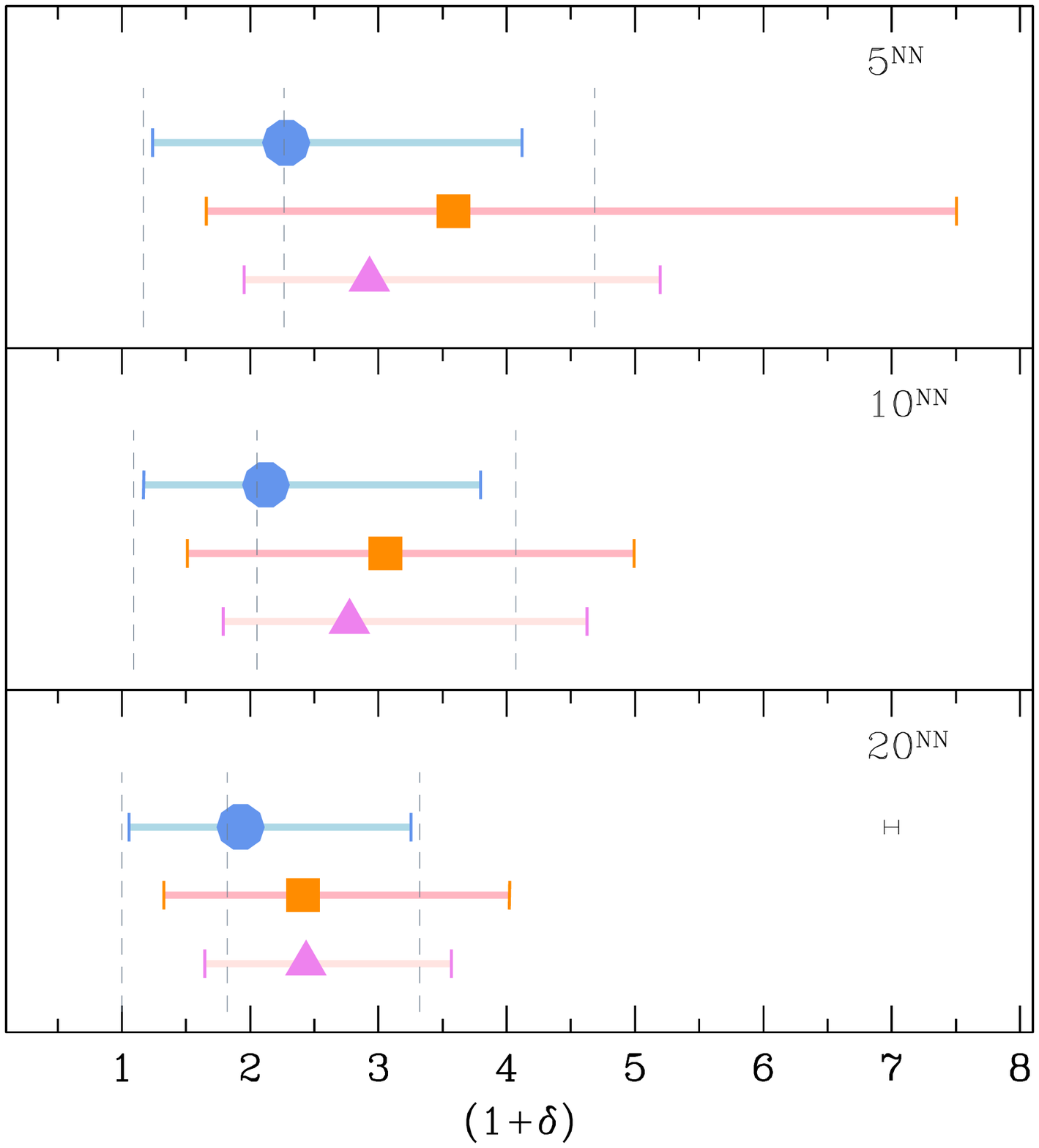}
\end{center} 
\caption{The environmental properties of the three classes of objects using
  different numbers of neighbours (5$^{th}$, 10$^{th}$, and 20$^{th}$ nearest
  neighbours, or NN).  In the left and right panels, we plot the distribution
  and statistics of the (1+$\delta_{5^{th}}$), (1+$\delta_{10^{th}}$), and
  (1+$\delta_{20^{th}}$) for each class of galaxies colour coded according to
  their classification (\psb\ with violet-colour coded, triangles; quiescent
  galaxies orange colour-coded, squares; and star-forming galaxies blue-colour
  coded, circles).  In both panels, the vertical dotted lines (gray colour)
  show the median and extreme quartiles for the 10k-parent sample. The black
  circles indicate the ratio of each class of galaxies with a particular
  density to the 10k-parent population computed in three equally-populated
  bins of galaxies.  In the right panel, the extreme quartiles of each
  distribution are used to show the range of density values in each
  population. The associated error bars obtained with resampling technique are
  plotted in the bottom, right panel.}
  \label{fig:env} \end{figure*}

There are two peculiarities of the zCOSMOS density field (cf. Fig.\,18 of
\citet{kno09}) that may affect our group-based zCOSMOS results. The first
peculiar feature in the group reconstruction is a lack of group galaxies in
the 10k sample compared to Mocks at $z\sim 0.55$.  The second peculiarity is
an excess of group galaxies at $z\sim 0.9$ that is not reproduced by any of
the Mock catalogues.  These peculiarities might be related to a significant
underdensity and overdensity observed at $z\sim 0.55$ and $\sim0.9$,
respectively, in the reconstructed three-dimensional density field by
\citet{kov09a}.  Finally, we note that at $z\sim 0.8$, the fraction of 10k
galaxies in groups is on average about 15\%, and never exceeds 30\%.

We now study the properties of \psb\ candidate galaxies in different
environments, and compare their properties with those of strong \ewoii\
emitters, quiescent galaxies, and the 10k-parent sample.
The environmental properties estimated using the 5$^{th}$, 10$^{th}$, and
20$^{th}$ nearest neighbour for zCOSMOS \psb\ galaxies and the comparison
samples are shown in Fig.\,\ref{fig:env}.  The typical physical scales for
which our analysis was performed, sampled by the three estimators at
$z\in[0.48-1.0]$, are $\sim 3.5$, 5.5, and 8~Mpc, respectively.

In the left panel of Fig.\,\ref{fig:env}, we plot the distribution of the
log(1+$\delta_{5^{th}}$), log(1+$\delta_{10^{th}}$), and
log(1+$\delta_{20^{th}}$) (from the left to the right panel) for \psb\
galaxies, and for quiescent and star-forming galaxies (from bottom to the top
panel). The distribution of the 10k-parent sample is overplotted for each
class of galaxy (dashed line).  The black circles indicate the respective
ratio of each class of galaxies to the 10k-parent population as a function of
the three environmental estimators computed in three equally-populated bins of
galaxies.

The bottom panel of Fig.\,\ref{fig:env} shows a deficit of
\psb\ galaxies with low values of overdensities (2\% with
log(1+$\delta$)$<0.1$) that increases to $\sim$ 5\% for higher
overdensities. Their distribution is shifted to higher overdensities
than that of the 10k-parent population. This trend is supported by the
low probability (at a significance level of more than $3\sigma$) that
the two distributions are drawn from the same 10k-parent
population. As expected, the fraction of quiescent galaxies also
increases at higher overdensity values, illustrating their preference
to reside in denser environments than the global population of
galaxies (Fig.\,\ref{fig:env}, middle panel). Using the spectroscopic
definition described in Sect.\,\ref{sec:spec}, the fraction of
quiescent galaxies doubles from 15\% in underdense regions
(log(1+$\delta$)$<0.1$) to 30\% in overdense regions ($>0.5$).  The
top panel of Fig.\,\ref{fig:env} shows the frequency of highly
star-forming galaxies relative to the total number of the 10k-parent
galaxies in different density regions.  As demonstrated in detail by
\citet{cuc09}, but see also \citet{iov09} for the complex interplay
between groups and galaxy properties, there is no significant
correlation between the star-forming galaxies in zCOSMOS and their
environment\footnote{The possible trend of a higher fraction of
  star-forming galaxies being in underdense environment when using the
  mass-weighted overdensities has been explored in \cite{cuc09}, but
  it disappears when using number densities (as employed here).}.  It
is useful to note that the definition of star-forming galaxies in
\citet{cuc09} does not differ significantly from the one adopted in
this work. A star-forming galaxy was defined in \citet{cuc09} to be an
object with a SFR higher than $\sim 16$\msunyr, calibrated using the
\oii\ luminosity following \citet{mou06}.

Globally, the distributions of both star-forming and quiescent galaxies have a
low probability (at $>3\sigma$) of being drawn from the same 10k-parent
population.

In the right panel of Fig.\,\ref{fig:env}, we plot the median values
of the three estimators for all the samples (\psb\ with violet-coded
triangles, quiescent with orange squares, and star-forming galaxies
with blue circle symbols). The first ($25^{th}$) and fourth
($75^{th}$) quartiles of each distribution are plotted to show the
range in density of each population.  By dashed, vertical lines, we
indicate the corresponding statistical values of the 10k-parent
catalogue.  The typical error bar for environmental quantities in our
galaxy samples is obtained by a resampling technique and plotted in
the bottom right panel of Fig.\,\ref{fig:env}. The representation of
median values and the distribution of the central quartiles confirm
also that quiescent and \psb\ galaxies reside in an overdense
environment with respect to 10k-parent galaxies. We note that this
environmental dependence is observed on different physical scales for
both galaxy types with a statistically significance of more than
3$\sigma$ level, and that in addition this dependence in quiescent
galaxies is stronger on progressively smaller physical scales. On the
other hand, star-forming galaxies do not appear to reside in
significantly different environments from the 10k-parent galaxies on
all physical scales (from 8~Mpc down to $\sim 2$~Mpc).

We are now able to investigate the impact of the environment probed on group
physical scales, on the properties of \psb\ galaxies. Using the optically
selected sample of group galaxies by \cite{kno09}, we measured and compared the
fraction of galaxies in each class identified as group members. In
computing these fractions, we accounted for the redshift dependence of the group
richness by using an effective richness (see for details Knobel et
al. 2009). 

Using the 10k catalogue of groups with a group defined to have either
two or more than three members, \psb\ galaxies do not appear to exist
preferentially in a group environment; the fraction of \psb\ galaxies
in groups is comparable to that of star-forming galaxies ($\sim 20$\%)
and lower than that of quiescent galaxies (using \mcor\ selection
criteria).  Using the more conservative criterion on mass
completeness, i.e., selecting only galaxies more massive than
\logm$>10.8$, we observe that the fractions of star-forming and \psb\
galaxies in groups increase, and become similar to that of quiescent
galaxies ($\sim 30$\%).

We note that the majority of both \psb\ and quiescent galaxies residing in
groups have an early-type morphology ($\sim 71$\% and $\sim 85$\%,
respectively), compared to $\sim 17$\% for the star-forming galaxies in
\smcor. However, for smaller numbers of objects, similar values are found for
\smst, which support these findings.

Based on these results, we conclude that although \psb\ galaxies are
found preferentially in overdense environments on typical scales of
clusters and above, they also exist in groups and in the field. Given
the results for the typical \ka\ galaxy environment using nearest
neighbours estimators and group catalogues, we found that on average
\ka\ galaxies prefer physical scales of $\sim 2- 8$~Mpc, i.e.,
overdense environments similar to that of quiescent galaxies. We also
identified \psb\ galaxies in groups (thus on physical scales smaller
than Mpc), although they tend to share a group environment more
similar to that of star-forming galaxies.  Given that during the short
interval of the \psb\ phase ($\approx 1$~Gyr), a galaxy cannot be
transported from an under- to an overdense environment, our results
imply that the $k+a$ spectral features are produced by a series of
mechanisms among the several proposed in the past and not necessarily
related only to environment.

Similar results were obtained by \citet{pog09} at $z=0.4-0.8$ using
the ESO Distant Cluster Survey, who robustly proved the higher
incidence of \ka\ galaxies in high density environments (in clusters
and in groups with a low fraction of \oii\ emitters) than in the
field, and pointed out that several different processes can produce a
\ka\ spectrum.  In contrast, \cite{yan08} in their study of DEEP2 data
found a higher fraction of \psb\ galaxies in field rather than group
environments at 1$\sigma$ significance level. We also note that
galaxies in our groups may have a wide range of physical properties.
This scatter in physical properties reflects the environmental
dependence of \psb\ galaxies \citep[e.g.,][]{pog06}. As shown in
detail by \citet{pog09}, \ka\ galaxies in groups with a low (or high)
fraction of \oii\ emitters have higher (lower) dependencies on
environment.  Finally, both merging or/and accretion events, and
cluster-related mechanisms such as ram-pressure gas stripping,
harassment, or strangulation, are efficient in suppressing
star-formation activity in a galaxy. Multiple possible origins of
$k+a$ galaxies have been proposed and we indeed confirm that a wide
range of origins is likely.
 
\subsection{Contribution to the mass assembly and spectral properties}
\label{sec:spectra}

The incidence of \psb\ galaxies in the general galaxy population varies
between authors, depending on the environment studied, redshifts, selection
methods, and properties of the various surveys.  In cluster environments and
at intermediate redshift, the fraction of \psb\ galaxies varies from high
values, e.g., $\sim 20-25$\% \citep{dre83, dre99, dre92, tran03, tran04,
tran07, pog99}, to intermediate values \citep[e.g., 10\%,][]{pog09}, to very
low values ($\sim 0.2$\% by \citet{bal99, yan08}) both in the field and
cluster environment at intermediate redshift.  At lower redshifts, \psb\
galaxies are rare \citep{zab96, blak04, got07} and reside mostly in the field.
It is clearly difficult to compare these \ka\ galaxy results for parent
samples selected from different environments and cosmic times and compiled
with different selection criteria. We therefore established our own internal
means of assessing the contribution of \ka\ galaxies to galaxy stellar mass
assembly and galaxy evolution in general.

We found a number density of \psb\ galaxies of $\sim 4.6\pm0.3 \times
10^{-5}$~Mpc$^{-3}$ and a mass density of $\sim 2.6\pm0.2 \times
10^{5}$~\msun~Mpc$^{-3}$ over the redshift range $0.48 < z < 1.2$.
The errors were estimated using resampling techniques: we estimate the
sample statistics $f_n$ using all $n$ galaxies, then remove one point
at a time to estimate $f_{{n-1},i}$. We found no statistically
significant evolution with redshift in either the number or mass
densities of \psb\ galaxies, but given the limited number of galaxies
in each redshift bin ($\sim$~10) a more solid conclusion will be drawn
at the completion of the survey.

To establish the impact of the \ka\ galaxy population on galaxy
evolution, we estimated the total mass flux that enters the red
sequence after a ($<1$~Gyr) quenching of the star-formation
activity. We followed the method of \citet{mar07} who introduced this
parameter in their study of SDSS/GALEX red and blue galaxies defined
with NUV-r colours, which was subsequently used in other works, e.g.,
by \citet{arn07} on a SWIRE-VVDS-CFHTLS data set. We defined the flow
in stellar mass by \psb\ galaxies that contribute to the red-sequence
mass assembly as

\begin{equation}\label{eq:massflux}
\dot{\rho}_{k+a \rightarrow RedSeq} =
  \frac{1}{t_{k+a}}\sum_{i=1}^{N_{k+a}} \frac{M^*_i}{w_i . 
  V_{max,i} } \,,
\end{equation}

\noindent where $M^*$ is the stellar mass in solar masses of \ka\
galaxies in the \smcor, $w$ is the selection function ($w_{\rm SSR}$
and $w_{\rm TSR}$ as introduced in Sect.\,\ref{sec:weight}), and
$V_{max}$ is the maximum volume in which each galaxy of a given
$I_{AB}$ magnitude is still observable. We assume that the \ka\
spectroscopic features are detectable for $t_{k+a}=0.35$~Gyr, or
1~Gyr.  Our \ka\ galaxies correspond to a mass flux entering the
red-sequence of $\dot{\rho}_{k+a \rightarrow RedSeq} = 7.8 \times
10^{-4} \pm 1.2$~\msun~Mpc$^{-3}$~yr$^{-1}$ (or a contribution of
$\sim 8$\% to the growth rate of the red-sequence estimated to be
d{\sl M}/dt $\sim 10^{-2}$~\msun~Mpc$^{-3}$~yr$^{-1}$ by \citet{poz09}
and assuming $t_{k+a}=0.35$~Gyr, or $\sim 3$\% with $t_{k+a}=1$~Gyr).
Of course, these values are in agreement with the assumption that
galaxies remain on the red-sequence and any process of re-juvenation
such as those proposed by \cite{has08}, are at work.

Results from the GDDS obtained by \citet{leb06} showed that galaxies
with strong \hdelta\ absorption (H$\delta$ strong galaxies, or HDSs)
decrease with cosmic time (from about 50\% at $z\sim1.2$ to a few
percent today).  However, they detected only one or two \psb\ galaxies
among the 25 HDSs (depending on how their error bars were treated). A
strong evolution in the number density of \psb\ galaxies was found by
\citet{wild08} (a factor of 200 higher at $z=0.7$ than in the local
Universe by \citet{qui04}), but on the basis of 5 \psb\ galaxies
selected only with a quite conservative definition in the VVDS-F02
catalogue.
Based on that sample, \citet{wild08} estimated that \psb\ galaxies can
account for $\sim 40$\% of the growth in the red sequence in the field
at $z=0.7$. Instead, \citet{delucia09} estimated that the \psb\ phase
is not the dominant channel that moves galaxies from the blue star
forming cloud to the red sequence in clusters at $z=0.5-0.8$.

\begin{figure}[ht!]
\begin{center}
\includegraphics[width=7cm,angle=-90]{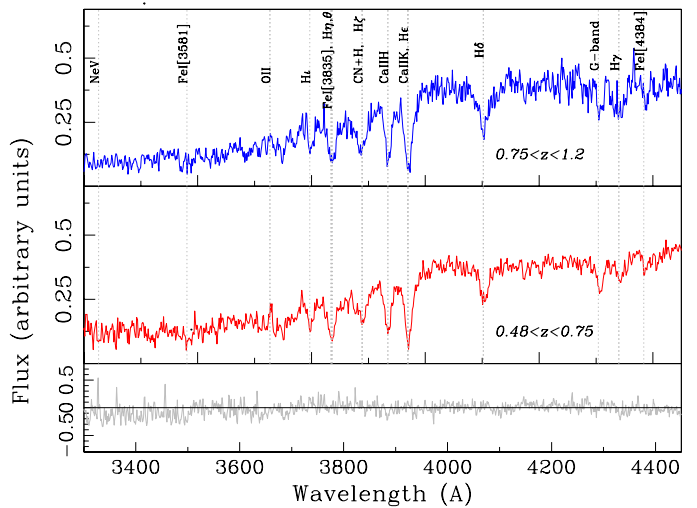}
\end{center}
\caption{Composite spectra of \psb\ galaxies at $z\in[0.75-1.2]$ and
$z\in[0.48-0.75]$ (and the non-statistically difference between them in the
bottom panel).}
\label{fig:combi} \end{figure}

We now investigate the possible evolution in the spectral properties of our
\psb\ galaxies. We use spectra of \psb\ galaxies (in \smcor) at high redshift
($z\in[0.75-1.2]$) and low redshift ($z\in[0.48-0.75]$).  In the three panels
of Fig.\,\ref{fig:combi}, we show from top to bottom the composite spectra of
\psb\ galaxies (in \smcor) at high redshift, low redshift, and the difference
between them. The redshift bins are chosen to include a comparable number of
\psb\ galaxies (40 spectra at low $z$, and 28 at high $z$), although the
spectra are weighted using our selection function before the combination. The
median redshifts of the two composite spectra are $z_{med}=0.659$ and
$z_{med}=0.895$ at low and high redshifts, respectively.  We shifted our
spectra to the galaxy rest-frame defined by their spectroscopic redshifts, and
normalized them to the common region around 4100--4700$\lambda$ sampled in each
spectrum. To each spectrum, we applied the weighting approach described in
Sect.\,\ref{sec:weight} obtaining a composite spectrum with a median
combination.

We do not observe any significant difference between the spectral
properties of \psb\ galaxies at $z_{med} = 0.659$ and $z_{med} =
0.895$. The continuum of the difference spectrum is centred on zero at
all wavelengths (apart from at very blue $\lambda$, where the
difference reaches 5\%). It is, however, true that there is not a very
large difference in redshift between these two spectra, so negligible
evolution given the uncertainty levels, may not be a surprising
result.

\section{Discussion}

A number of studies have explored the physical properties and the
relative importance of \ka\ galaxies, i.e., galaxies with strong
Balmer absorption lines (which are indicative of an intense
star-formation epoch in the past billion years) and an absence of
emission lines (a signature of lack of ongoing star formation).
Even considering the different numbers cited in the literature about
the incidence of \ka\ galaxies among the entire galaxy population,
and their apparent or true evolution in number density (which has not
been confirmed in this work), they represent a minor constituent of
the galaxy population in the Universe in the past 8~Gyr.
These systems are physically interpreted as galaxies that have had
their star formation abruptly quenched in the past billion years and
subsequently migrated onto the red sequence.  Thus, the importance of
this galaxy population depends on its role in creating the red
sequence, or in stellar mass assembly in general. Other important
issues are the mechanism/s responsible for the abrupt quenching of
star formation in their progenitors and the true nature of their
descendants.

We have completed an analysis of high quality spectral data provided by
 zCOSMOS and ancillary data available for the COSMOS field of the first
 mass-selected sample of 69 \ka\ galaxies in a wide range of environments at
 intermediate redshift ($0.48<z<1.2$).

Alternative explanations of why \ka\ galaxies appear to be transient objects
have also been considered. Using 1.4~GHz radio and X-ray data, we have
confirmed that our \ka\ galaxies are unlikely to be strongly star-forming
galaxies affected significantly by dust extinction and only a fraction of them
may be obscured AGNs. We have also considered another concern that Balmer
line strengths might be generated by some other hot population than the
main-sequence turn-off stars, e.g., blue stragglers \citep[e.g.,][]{ros85,
ros94} or blue horizontal branch stars \citep[e.g.,][]{bur84, mt00,
tw05}. 

In this work, we adopted a conventional definition in selecting \ka\
galaxies.  Although \halpha\ is a robust indicator of star formation,
it falls out of the optical window at intermediate redshifts and bluer
diagnostics are used in assessing the star-formation activity of the
\psb\ galaxy population.  \citet{yan06} reported on the incompleteness
introduced in using \oii\ to identify \ka\ galaxies, a criterion that
could unfairly eliminate \psb\ galaxies with AGN activity.  This is
because the \oii\ emission line is sensitive to both the star
formation and AGN activity in a galaxy, and by including galaxies on
the basis of the lack of an \oii\ emission line, we are a priori also
removing those with an AGN contribution. However, we followed the
classical approach of using the \oii\ diagnostic instead of the
\hbeta\ line proposed by \citet{yan08}, because \hbeta\ is a far
weaker line than \oii\ which can increase the uncertainty in e.g., the
stellar continuum fit, and is affected by both emission and
absorption. Thus, in this paper we adopt the conventional definition
that involves both the \oii\ emission line and \hdeltaA, to maximize
the comparison with the majority of other works. Any impact of the
\ka\ galaxy population in a general context of galaxy evolution may be
slightly greater than quantified here. Our incompleteness should
however be minimal because the duty cycle ($\sim 10^8$~yr, e.g.,
Condon et al. 1992) of a possible AGN in our \ka\ galaxies represents
a short period in the lifetime of a galaxy.

We note that any comparison with external results is possible only
qualitatively, while a more objective analysis is feasible with
internal comparisons within a given catalogue as pointed out by
several authors. We followed this approach in our work by objectively
comparing our results on \ka\ galaxies with star-forming and quiescent
galaxy mass-selected samples defined using spectroscopic criteria and
with 10k-parent mass-selected galaxies, all extracted from the zCOSMOS
bright program. The results obtained in this work are robust and
corrected for effects of incompleteness.

We explored the physical properties of \ka\ galaxies in terms of
stellar mass, mass-to-light ratio, colour, morphology, and environment.
We found that our \ka\ galaxies are as massive as quiescent galaxies, but have
relatively lower mass-to-light ratio and bluer colours, which ensure that they
are potential contributors to the red sequence and ideal occupants of the
green valley, i.e., the underpopulated region in-between the distribution of
star-forming/blue galaxies and quiescent/red galaxies.  We confirmed that
morphologically zCOSMOS \ka\ galaxies are a heterogeneous population of bulge-
and disc-dominated galaxies, which become predominantly more bulge-dominated at
increasing stellar mass. Despite the high incidence of asymmetries indicative
of recent interaction events, our morphological results have demonstrated that not
all progenitors of our \ka\ galaxies are spheroidals as expected if a
major merger had destroyed the disc.  

Our finding that 50\% of our \ka\ population are passive spiral
galaxies (but depending on stellar mass) and the knowledge that
in distant clusters, \ka\ galaxies are found to have predominantly
spiral morphologies \citep[e.g., by ][]{dre99,pog99} is consistent
with the timescale for morphological transformation being longer than
that for the transformation in spectral properties. However, because
of their high stellar masses and the large spread in their colours in
the green valley, star formation in at least a fraction of \ka\
galaxies may be {\it periodically} triggered by the rapid accretion of
outer fresh material (as e.g., proposed by the re-juvenation
hypothesis of \cite{has08}).  Given that \ka\ galaxies are found in a
wide range of environments (from field to cluster scales), the
mechanisms quenching star formation in \ka\ galaxy progenitors and
their descendants might both be diverse.  Summarizing, the results
obtained in the present work provides good support to the common
interpretation for \ka\ galaxies that they may represent a transient
phase between star-forming and a quiescent stage of star-formation
activity, but their subsequent evolution may be to a wide range of
galaxy classes by means of many different evolutionary paths.

It is important to determine the way in which star formation is
controlled and truncated in a galaxy. Star formation can experience a
gradual decrease as would be consistent for example with the smooth
decline in redshift ($z < 1$) of the star formation rate, or star
formation can be suddenly and violently extinguished. Given that the
spectra of \ka\ galaxies imply that they have had their star formation
truncated suddenly in the recent past, measuring their number and mass
density can help us to quantify the significance of such phenomenon in
suppressing the global star formation rate. \Ka\ galaxies have
abruptly quenched their star formation activity on a timescale shorter
than a billion years, therefore their contribution represents a lower
limit to the prevalence of such sudden quenching.  The decline in the
star formation activity in galaxies is understood to have occurred
gradually with cosmic time in the past 8~Gyr, although its dependence
on the stellar mass remains controversial, e.g., compare \citet{zhe07}
and \citet{dro08}.
We computed the contribution of \ka\ galaxies to the mass assembly of
the red-sequence in the conservative hypothesis that they remain on
the red-sequence. This contribution to the growth rate of the
red-sequence has been found to be in the range between 3\% and 8\%
depending on the assumed timescale of the characteristic detectability
of \ka\ spectroscopic features.  Considering a number of possible
underestimation effects such as e.g., the removal of \ka\ galaxies
with obscured AGNs and the non-detection of \ka\ objects because of
the insufficient signal-to-noise ratio of the spectra, we conclude
that although the \ka\ galaxy population is not the primary
contributor to the red sequence, it might not be negligible in the
general context of galaxy evolution. Although sudden truncations of
the star formation activity appear not to be more prevalent than
smoother declines in the redshift interval $0.48<z<1.2$, they are
probably still important. In addition, a lack of sufficient temporal
resolution in the star formation history cannot prevent us from
speculating that the star formation might occur in episodes of
moderate intensity separated by short quiescent phases as observed in
nearby dwarf galaxies in which the resolved stellar populations can be
studied in great detail \citep{eline09}. In this scenario, the \ka\
spectral features are common in the life of each galaxy, but observed
only in a small fraction of the entire galaxy population given their
short lifetime.

\section{Conclusions}

We have explored the physical properties of the first mass-selected sample of
\ka\ galaxies at intermediate redshift ($z=0.48-1.2$) in a wide range of
environments. We selected this galaxy population, interpreted to represent a
link between star-forming/blue galaxies and quiescent/red galaxies, from the
spectroscopic program zCOSMOS related to the largest multiwavelength survey
completed to date, COSMOS.
\medskip

Our results can be summarized as follows:

\begin{enumerate}
\item{\Ka\ galaxies occupy the brightest tail of the luminosity
    distribution. They can be as massive as typical quiescent galaxies
    and populate the so-called ``green valley'' in the colour versus
    luminosity (or stellar mass) distribution.}

\item{A fraction ($<8$\%) of these galaxies have radio and/or X-ray
  counterparts implying an upper limit to the SFR of $\sim 8$\msunyr. These
  results suggest a possible contribution from obscured AGNs.}  

\item{\Ka\ galaxies morphologically represent a heterogeneous
    population with a similar incidence of bulge-dominated and disky
    galaxies, independent of the strength of their \hdelta\ absorption
    line, but dependent on stellar mass in a way reminiscent of the
    well-known mass-morphology relation (i.e., the most massive \psb\
    galaxies are predominantly bulge-dominated systems).  We reported
    a high incidence of asymmetries in HST/ACS images, supporting the
    hypothesis that these galaxies have experienced interaction/merger
    processes in their recent past, although we are unable to
    ascertain whether these processes are entirely responsible for the
    quenching of the galaxy star-formation activity.}

\item{\Ka\ galaxies reside preferentially in quite rich environments
    on physical scales of $\sim 2-8$~Mpc and have morphological
    early-to-late type ratios in groups that are similar to those of
    quiescent galaxies, although we also found \ka\ galaxies in
    underdense regions. We concluded that several different
    mechanisms, not necessarily related to a specific environment,
    quench star-formation activity in galaxies on short timescales
    ($<1$~Gyr).}

\item{We measured a number density of \ka\ galaxies of $\sim 4.6\pm0.3
    \times 10^{-5}$~Mpc$^{-3}$ and a mass density of $\sim 2.6\pm0.2
    \times 10^{5}$~\msun~Mpc$^{-3}$ over the redshift range $0.48 < z
    < 1.2$.  With the current data set, we did not find statistical evolution
    in either any of the aforementioned quantities in the redshift
    interval explored, nor by comparison with the local values of
    \citet{qui04}. We were also unable to determine any evolution in
    the spectral properties between $0.48 < z \le 0.75$ and $0.75 < z
    \le 1.2$.}

\item{We studied the contribution of \ka\ galaxies to the mass
    assembly of the red-sequence. This contribution can be as high as
    $\sim 8$\% ($\dot{\rho}_{k+a \rightarrow RedSeq} = 7.8 \times
    10^{-4} \pm 1.2$~\msun~Mpc$^{-3}$~yr$^{-1}$) given a growth rate
    in the red-sequence of d{\sl M}/dt $\sim
    10^{-2}$~\msun~Mpc$^{-3}$~yr$^{-1}$ (as estimated by Pozzetti et
    al. 2009 for the same data) and assuming a detectability of \ka\
    spectroscopic features of $t_{k+a}=0.35$~Gyr.
  Although \ka\ galaxies may not be the primary source of growth to
  the red sequence, they are likely to be important to galaxy
  evolution. Furthermore, sudden truncations in star formation {\it do
    not appear to be} the primary means of quenching the star
  formation activity compared to more gradual means, but appear
  nevertheless to be important to the decline in the global star
  formation activity.
  On the other hand, lacking sufficient temporal resolution, we cannot
  exclude the scenario where the sudden quenching of star formation
  and the regime in which the star formation occurs in episodes of
  moderate intensity separated by short quiescent phases are the rule
  in general galaxy evolution. It is worth noting indeed that in
  nearby dwarf galaxies when the resolved stellar populations are
  studied in great detail, the star formation history of these
  galaxies appears to consist of repeated periods of star formation
  \citep{eline09}. More solid results on this and other subjects will
  be obtained using the larger 20k zCOSMOS sample.}

\end{enumerate}

\section{Acknowledgments}

This work has been supported in part by the grant ASI/COFIS/WP3110
I/026/07/0. We thank the referee for providing constructive comments.

\bibliographystyle{aa} \bibliography{biblio_psb}

\end{document}